\documentclass[epj]{svjour}

\usepackage{graphics}
\usepackage{epsfig}

\begin{document}

\title{Salt Modulated Structure of Polyelectrolyte-Macroion Complex Fibers}

\author{Hoda Boroudjerdi\inst{1} \and Ali Naji\inst{1,2} \and Roland R. Netz\inst{1}}

\institute{Department of Physics, Technical University of Munich,
Garching 85748, Germany \and Department of Applied Mathematics and Theoretical Physics, 
Centre for Mathematical Sciences, University of Cambridge, Cambridge CB3 0WA, United Kingdom}

\date{Received: date / Revised version: date}


\abstract{
The structure and stability of strongly charged complex fibers, formed by complexation of
a single long semi-flexible poly\-electro\-lyte chain
and  many oppositely charged spherical macroions, are investigated numerically at the ground-state
level using  a chain-sphere cell model. The model takes into account chain elasticity as well as electrostatic 
interactions between charged spheres and chain segments.  Using a numerical optimization method
based on a periodically repeated unit cell, 
we obtain fiber configurations that minimize the total energy.  
The optimal fiber configurations exhibit a variety of helical structures for the arrangement 
of macroions including zig-zag, solenoidal and
beads-on-a-string patterns. These structures are determined  by a competition between attraction between spheres
and the  poly\-electro\-lyte chain (which favors chain wrapping around the spheres), chain bending 
and electrostatic repulsion between chain segments (which favor 
unwrapping of the chain), and the interactions between neighboring sphere-chain complexes which can be attractive 
or repulsive depending on the system parameters such as medium salt concentration,
macroion charge and chain length per macroion (linker size). At about physiological salt concentration, dense zig-zag patterns
 are found to be energetically most stable when parameters appropriate for the
 DNA-histone system in a chromatin fiber are adopted.  
In fact, the predicted fiber diameter in this regime is found to be around 30 nanometers, which appears to agree with the thickness
observed in {\em in vitro} experiments on chromatin. We also find a macroion (histone) density of 
5-6 per 11~nm which agrees with the zig-zag or cross-linker models of chromatin.  Since our study deals primarily 
with a generic chain-sphere model, these findings suggest that 
structures similar to those found for chromatin should also be observable for poly\-electro\-lyte-macroion complexes formed
in solutions of DNA and synthetic nano-colloids of opposite charge. 
In the ensemble where the mean linear density of spheres on the chain is fixed,
the present model predicts a phase separation at intermediate salt concentrations into a densely packed
complex phase and a dilute phase. 
} 

\maketitle


\section{\label{sec:intro} Introduction}

Recent years have witnessed a growing interest in the structure and phase behavior of large charged 
polymer-macroion clusters, for instance, a long poly\-electro\-lyte (PE) chain is complexed with 
many oppositely charged spheres \cite{Bielinska,Dubin88,Tsuboi,Gana,Strauss,Xia,Haronska,Gittins,Caruso,Li,Dubin90,Dubin1,Linse1,Jonsson1,Skepoe}.
Perhaps, the most striking example is realized  in biology, where a huge 
fiber is formed by complexation of a long (negatively charged) DNA chain with (positively
charged) histone proteins, giving rise
to the prominent  {\em chromatin fiber} \cite{The_cell,Schiessel_chromatin,Schiessel_rev,Kornberg,Luger1,Widom,Yao,Holde}. 
The fundamental unit of chromatin is known as the {\em nucleosome} 
consisting of about 200 base pairs (about 68~nm) of DNA  associated with a 
cylindrical-wedge-shaped histone octamer of diameter about 7~nm and mean height 5.5~nm. 
The nucleosome has two main parts: i) the {\em nucleosome  core particle}
(or the {\em core particle}) comprising 146 base pairs of DNA wrapped in nearly a 1-and-3/4  left-handed 
helical turn around the histone octamer, and ii) the {\em linker DNA}, which connects adjacent 
core particles to one another. 
There is an additional component known as linker histone H1 (or its variant H5), 
which binds to the DNA and the histone octamer in such a way that it brings the entering and exiting strands 
of DNA together along a short distance, forming a stem-like  structure.  The role of linker histone will be 
discussed further below.

{\em In vitro} experiments have revealed  striking salt-induced conformational changes for the chromatin  fiber 
(see, e.g., \cite{The_cell,Thoma,Allan,Gerchman,Woodcock,Bednar,Horowitz}): 
At very small salt concentration, the fiber structure is swollen, displaying an open {\em  beads-on-a-string}
pattern in which  individual core particles become highly separated from each other; the resultant fiber is known
as the {\em 10~nm fiber} because of the nearly 10~nm hard-core diameter of the core particle  (nearly equal to the combined
diameter of the core histone octamer and twice that of the DNA wrapped around it).
As the salt concentration increases, 
the fiber becomes more and more folded and finally within the physiologically relevant regime (around 
100~mM NaCl), it exhibits a thick and dense fiber of diameter about 30~nm, known as the
{\em 30~nm fiber}. This salt-dependent behavior indicates that predominant electrostatic mechanisms are involved and
influence the organization of the chromatin fiber. 

The precise arrangement of nucleosome core particles in the 30~nm fiber has been under 
debate for a long time \cite{Schiessel_chromatin,Schiessel_rev,Thoma,Woodcock,Bednar,Horowitz,Leuba,Finch,Worcel,Cui,Katritch,Wedemann,Schiessel_EPL,Mozziconacci2,Kulic04,Woodcock_rev,Richmond05,Richmond04,Robinson2006,Robinson2005,Langowski,Diesinger,Beard,Schlick06,vanHolde,Mozziconacci,Depken,Schiessel2009,Kruithof,Routh}. 
A large number of models have been proposed for the structure of the 30~nm fiber which are  more or less
consistent with the experimental observations.  
For instance, several studies \cite{Woodcock,Bednar,Horowitz,Leuba} appear to support a zig-zag or cross-linker pattern, where successive 
nucleosomes are at opposite sides of the fiber and the linker DNA  joining them bridges 
across the helical core of the fiber.
Among other models supported by recent experiments are the solenoidal model \cite{Thoma,Finch}  
(in which successive nucleosomes are juxtaposed as nearest neighbors with the linker DNA bent in between them),  
the twisted ribbon model \cite{Worcel},  as well as the more general class of ribbon models \cite{Depken,Schiessel2009} 
(see Refs. \cite{Schiessel_rev,Woodcock_rev,vanHolde,Schiessel2009} for a more comprehensive reference list and discussion of various chromatin models).

Our goal in this work is to address structural properties of generic chain-sphere complex fibers comprising 
a long semi-flexible charged PE chain and many oppositely charged spheres. 
We adopt a simple chain-sphere cell-model approach \cite{Kunze2,hoda1} 
that allows for a systematic description of the polymer conformation in an infinite fiber. 
The chain conformation and the fiber structure are determined 
by a competition between  chain elasticity, electrostatic interactions (among chain segments and 
charged spheres across a salt solution) as well as geometrical constraints. We employ a numerical optimization method
to find configurations minimizing the total fiber Hamiltonian with respect to 
the whole conformation of the complexed chain and the position of macroions. This scheme is known as the
 {\em ground-state-dominance approximation} which is valid for strongly-coupled
charged complexes (such as DNA-histone complexes) that exhibit large PE adsorption energy and relatively 
small thermal fluctuations \cite{Kunze2,hoda1}. 
The organization of the chain on individual macroions  
and hence throughout the fiber is {\em not} externally imposed, but rather obtained as a result of 
the interplay between inter- and intra-fiber interactions.
We demonstrate that the resultant optimal fiber structures exhibit a variety of helical 
structures including, for instance, beads-on-a-string, zig-zag, and solenoidal patterns depending on a few 
basic system parameters, namely, the salt concentration, the macroion charge valency, and the chain length per macroion (or the linker length).
Other parameters such as the chain persistence length and its charge density as well as the sphere diameter
will be fixed; here we choose the parameter values appropriate for the DNA-histone system throughout 
this study. We should emphasize, however, that the present model is not meant to address the specific problem of the chromatin fiber
(as no attempt is made in order to incorporate explicitly the specific structural details of the actual chromatin fiber such as the linker histone, core histone tails, 
specific binding effects or the precise shape of the histone octamer \cite{The_cell,Schiessel_rev}), but we 
rather focus on the more general aspects of complex fiber formation in PE-macroion systems. The results are
 thus also relevant  
for  mixtures of DNA and oppositely charged nano-colloids or synthetic spheres.

We show that the transition between the predicted helical patterns is dictated by the wrapping-dewrapping behavior of the chain as
the system parameters vary. For instance, at low salt concentration, the dominant self-repulsion of chain
segments leads to dewrapping of the chain from macroions, generating an expanded 
fiber. In the physiological regime, zig-zag patterns are found to be energetically 
most stable, because  the chain 
is wrapped around macroions  in about a 1-and-3/4 turn for moderate sphere charge \cite{Kunze2}, 
which turns out to be
about the same degree of DNA-wrapping in the nucleosome core particles.  
At elevated salt (beyond 100mM monovalent salt), the highly wrapped state of the chain leads to very compact solenoidal 
structures for the fiber. The predicted structures will be summarized by presenting a 
two-angle structural diagram in analogy with the two-angle models in Refs. \cite{Schiessel_chromatin,Woodcock}. 
An important result follows from our model when the chain (linker) length per macroion is treated as a free parameter. 
It is found that the binding energy (per macroion) takes a  non-convex shape at intermediate salt concentrations 
(when plotted as a function of the chain length per macroion), with a global minimum at a high sphere density,
indicating a  gas-liquid-type ``phase coexistence" along the fiber:
Part of the PE chain forms a dense complex fiber with macroions phase-separating from 
a dilute phase along the fiber consisting only of uncomplexed chain. 

In the aforementioned optimization model, 
the wrapping structure of the chain around individual macroions is not fixed
and can vary  so as to minimize the effective Hamiltonian of the fiber (unconstrained optimization). 
This can be thought of as a simple model mimicking linker-histone-depleted chromatin \cite{Thoma}. 
The 30~nm fiber obtained within the unconstrained optimization model shows a rather
small density (number per projected unit length) of macroions along the fiber, i.e., about one macroion per 11~nm, 
which is comparable with the nucleosome density at relatively small salt concentration (where the chromatin 
still has a 30~nm diameter \cite{Bednar}), but not with the density of about 6 histones 
per 11~nm \cite{Schiessel_rev,Thoma,Gerchman,Woodcock,Bednar,Horowitz} observed in the physiological salt regime.  
 

\begin{figure*}[t]
\begin{center}
\includegraphics[angle=0,width=14.cm]{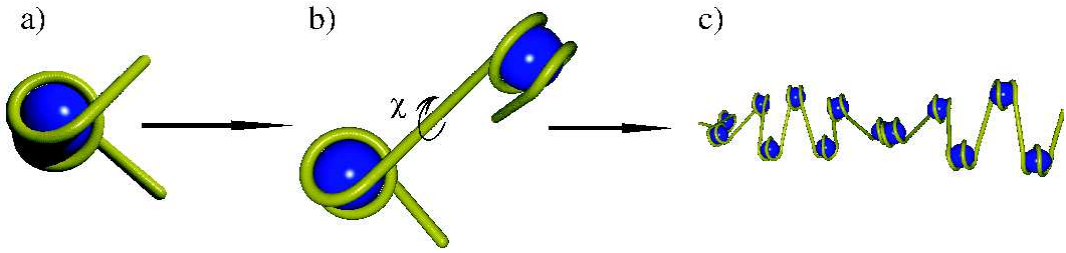}
\smallskip
\caption{\label{fig:model}Cell model for a complex fiber: a) Each unit cell consists of a single PE-macroion complex the precise
configuration of which is obtained from a numerical minimization method. b) Two adjacent unit cells 
(which are identical in configuration) are linked together such that they share a common tangent vector 
at the unit cells junction (see text). c) By repeating this
procedure the whole complex fiber is constructed from a main unit cell. Within the full minimization scheme, 
all degrees of freedom associated with a single unit cell (that is the chain beads locations as well as the boundary
rotational angle $\chi$) are varied  in order to find the optimal configuration minimizing the full effective Hamiltonian of the 
fiber.}
\end{center}
\end{figure*}

As noted above, the linker histone H1 ``glues" together the entering and exiting strands of the DNA
and also  stabilizes (``seals off") 
 the internal structure of the DNA in the nucleosome core particle. It plays a key role in the compaction of the chromatin into a 30~nm fiber \cite{Thoma}. 
 In order to mimic this kind of effect in a very simple way within the present model, we 
 also employ a variant of the optimization scheme, the so-called {\em constrained  optimization}, 
in which the core complex structure is fixed according to the optimal structure of an isolated complex 
 (with a chain length of 146 DNA base pairs); then only the relative orientation angle of adjacent  
core particle complexes can vary in order to find the optimal fiber structure. This leads to a number of significant changes.  
Most notably, the entering and exiting strands of the chain tend to form a cross pattern 
at physiological salt (producing
a small entry-exit angle as compared with the unconstrained model), which resembles the 
stem-like pattern proposed for H1-stabilized nucleosome  \cite{Thoma,Woodcock,Bednar}. 
As a result, the resultant optimal fiber is found to have a more compact zig-zag pattern with a histone
density of  about 5-6 histones per 11~nm in agreement with Refs.  \cite{Thoma,Gerchman,Woodcock,Bednar,Horowitz}. 

The organization of this paper is as follows: In Section \ref{sec:theory}, we describe the theoretical model and
the numerical method used to study infinite charged complex fibers. In Section \ref{sec:fiber_salt}, the salt-induced
structural changes of the fiber are studied within the unconstrained optimization model. The constrained optimization model
is analyzed in the following
Section  \ref{sec:chromatin_constrained}. We then show in Section  \ref{sec:length_var} how the fiber structure is 
influenced by variation of the linker length and discuss different thermodynamic ensembles and the possibility of
phase separation along the chain. 



\section{\label{sec:theory} Theoretical model: Cell-model approach}

The model used here for the study of complex fibers is based upon the chain-sphere model  \cite{Kunze2,hoda1}, 
which consists of a single semi-flexible PE chain complexed with an 
oppositely charged sphere (macroion).  Here we use this model to 
describe an infinite complex fiber made from complexation of a {\em single} long PE chain with many macroions. 
We restrict the discussion to strongly coupled complexes and thus employ the ground-state-dominance approximation corresponding to strong adsorption of the PE chain on each sphere \cite{Kunze2,hoda1,phys_rep}. 

The main idea here is to apply a {\em cell-model} approach by describing the complex fiber 
in terms of identical chain-sphere {\em unit cells} linked together via the same PE strand. 
In analogy with the chromatin fiber, one can 
thus think of each unit cell as consisting of  two main parts: i) the {\em core-particle complex} (macroion sphere complexed with
a chain segment), and ii) the {\em linker chain}
connecting adjacent core particles as shown in Fig. \ref{fig:model}. 

In our numerical analysis, the complex fiber is constructed by replicating one unit cell (referred to as the 
main unit cell) infinitely many times such that the {\em connectivity and smoothness} of the linker chain 
is preserved. We begin the minimization with a given initial condition, typically, the one in which the structure of the
core particle complex in each unit cell corresponds to the energetically optimized configuration in an {\em isolated}
chain-sphere complex with the chain length associated with each core particle being taken to be equivalent to 146 DNA base pairs. 
This procedure for a single isolated complex has been studied thoroughly in our previous works \cite{Kunze2,hoda1,thesis}. 
We then connect an arbitrary length of linker chain
and the interaction Hamiltonian of the {\em whole} fiber is minimized using a recursive numerical method 
to find the optimal fiber structure without any further constraints (see below).  This follows simply 
by focusing on a single unit cell. Although this prescribes a model consisting of 
identical unit  cells, the optimized fiber structures  are by no means trivial in the way the unit cells or core particles 
are arranged. They display a rich structural phase diagram  of helical patterns.  
Note that these helical patterns may or may not exhibit periodic structures (i.e., structures with 
discrete translational symmetry). In fact, the energetically optimal
structures are merely quasi-periodic, i.e., they do not show exact (rational) periodicity but
a distribution of periodicities with a sharp peak around a mean value of periodicity
(see Ref.  \cite{thesis} for details). 
We also note that electrostatic interactions between all unit cells are fully taken into account, and thus on this level,  the only
approximation of the present cell model is that structural symmetry breaking between 
the unit cells is neglected.

\subsection{Main unit cell}

Each unit cell consists of a piece of PE chain of length $L_c$ (or an equivalent of 
$N_{\mathrm{bp}}=L_c/(0.34~{\mathrm{nm}})$ DNA base pairs each of length $0.34$~nm) with linear charge density of 
$-\tau$ (in units of the elementary charge $e$) and 
bare (unscreened) mechanical persistence length of $\ell_{\mathrm{p}}$. The PE chain interacts with  a uniformly
charged sphere of radius $R_{\mathrm{s}}$ and charge valency $Z$ in a salt
solution of concentration $C_s$. Although the present model is quite general, 
in making explicit calculations, 
we set the parameters appropriate for the DNA-histone system by choosing $\tau=5.88$~nm$^{-1}$
(maximum dissociation  of two elementary charges per base pair \cite{note_tau_max}), 
$\ell_{\mathrm{p}} = 30$~nm \cite{Kunze2,Rief99,Frontali}
and  $R_{\mathrm{s}}=5$~nm, while $L_c$, $Z$ and $C_s$ are varied.
Note that the sphere radius of $R_{\mathrm{s}}=5$~nm represents
the closest approach of DNA of radius 1~nm on a histone octamer of mean radius of about 4~nm.
 As in previous studies \cite{Kunze2,hoda1},
 we employ a discretization scheme 
in order to parametrize the chain conformation as required for numerical analysis. 
The chain is discretized using $N+1$ discretization points or {\em beads}  of  
valency $q=\tau L_c/(N+1)$ located at positions 
$\{{\mathbf r}^0_i\}$ (with $i=0, 1, \dots, N$ and the super-index 0 denoting the {\em main unit cell}). 
This amounts  to $N$ rigid chain subunits within a unit cell each described by a 
{\em bond vector} ${\mathbf u}^0_i = {\mathbf r}^0_i-{\mathbf r}^0_{i-1}$ of 
length  $|{\mathbf u}^0_{i}| = \Delta$ and  two polar and azimuthal angles
$\theta_i$ and $\phi_i$ with respect to a fixed orthogonal reference \cite{Kunze2,hoda1}
($N$ has in general no connection with the number of actual monomers). 
The location of the central macroion ${\mathbf R}^0$  in the main unit cell is chosen
as the origin, i.e.,  ${\mathbf R}^0={\mathbf 0}$. 
 We conventionally discretize each natural DNA base pair into two discrete subunits; therefore 
 $N=2N_{\mathrm{bp}}$ and we have $q=1$ (i.e., one elementary charge per subunit or two per base pair) 
 for large $N$ when DNA linear charge density
 is adopted.

\subsection{Unit cell replication}
\label{subsec:chromatin_periodic}

Let us assume that the structure of the core-particle complex within the main unit cell is given  
(from a minimization scheme as discussed later). We construct the complex fiber by attaching 
an {\em image} of the main unit cell to it such that the linker chain {\em tangentially} connects the two unit cells
with no kinks at the junction (see Fig. \ref{fig:model}); i.e., the last  chain bond  vector, ${\mathbf u}^0_N$, 
 in the main unit cell  is chosen as the first bond vector, ${\mathbf u}^1_{1}$, 
 of the first immediate image cell (labeled by super-index 1) and so on for all other image cells
 in a recursive manner. 
 
Formally, this replication procedure is established by a combined translation
and rotation transformation.  
In general,  the position of the $i$-th chain bead and that of the sphere center 
in the $k$-th unit cell (labeled by super-index $k$), i.e., ${\mathbf r}^k_i$ and ${\mathbf R}^k$ respectively, are 
obtained as
\begin{eqnarray}
{\mathbf r}^k_i &=& {\mathbf E}\cdot ({\mathbf r}^{k-1}_i-{\mathbf r}^{k-1}_0) + {\mathbf r}^{k-1}_{N-1}\\
\nonumber\\
{\mathbf R}^k &=& {\mathbf E}\cdot ({\mathbf R}^{k-1}-{\mathbf r}^{k-1}_0) + {\mathbf r}^{k-1}_{N-1}, 
\end{eqnarray}
from the given configurations in the $(k-1)$-th unit cell (labeled by super-index $k-1$). 
Here  ${\mathbf E}$ is the full rotation matrix around the 
{\em common point} or junction ${\mathbf r}^{k-1}_{N-1}={\mathbf r}^k_0$ 
of the two adjacent unit cells, which rotates the bond vector ${\mathbf u}^k_{1}$ (and in fact the
whole $k$-th unit cell as a rigid body) such that ${\mathbf u}^k_{1}$
falls onto ${\mathbf u}^{k-1}_N$ to fulfill the smoothness of the chain at the junction.  We then 
allow the image unit cell $k$ to rotate around this common bond vector by an arbitrary (azimuthal) 
angle $\chi$,  see Fig. \ref{fig:model} for an illustration. This angle is indeed a rotational degree of freedom for each unit cell as a whole 
in addition to other intra-cell degrees of freedom (bead and sphere positions) that is varied in order to minimize the
fiber Hamiltonian (see Section \ref{subsec:chromatin_fiber_min_method} below).

The rotation matrix ${\mathbf E}$  is  obtained as
\begin{equation}
	{\mathbf E} = {\mathbf B} \, {\mathbf X} \, {\mathbf A}
\end{equation}
with the matrices ${\mathbf A}$, ${\mathbf X}$ and ${\mathbf B}$ defined as follows
\begin{eqnarray}
{\mathbf A} &=&
\left( \begin{array}{ccc}
-\sin\phi_1 & \cos\phi_1 & 0 \\
-\cos\phi_1\cos\theta_1 & -\sin\phi_1\cos\theta_1 & \sin\theta_1 \\
\sin\theta_1\cos\phi_1 & \sin\phi_1\sin\theta_1 & \cos\theta_1 \end{array} \right), 
\\
{\mathbf X} &=&
\left( \begin{array}{ccc}
\cos\chi & -\sin\chi & 0 \\
\sin\chi & \cos\chi & 0 \\
0 & 0 & 1 \end{array} \right), 
\\
{\mathbf B} &=&
\left( \begin{array}{ccc}
-\sin\phi_N & -\cos\phi_N\cos\theta_N & \sin\theta_N\cos\phi_N \\
\cos\phi_N & -\sin\phi_N\cos\theta_N & \sin\phi_N\sin\theta_N \\
0 & \sin\theta_N & \cos\theta_N \end{array} \right), 
\end{eqnarray}
where $\theta_1$ and $\phi_1$ are the polar and azimuthal angles of ${\mathbf u}^0_1$ and  $\theta_{N}$ and $\phi_{N}$
are the polar and azimuthal angles of ${\mathbf u}^0_{N}$. Note that the matrix ${\mathbf E}$ is only  a function of
the angles $\{\chi, \theta_1, \phi_1, \theta_{N}, \phi_{N}\}$. 

\subsection{Intra-fiber interactions}
\label{subsec:chromatin_fiber_interaction}
 
Because the infinite complex fiber is obtained by replicating identical unit cells, 
one can focus on the effective Hamiltonian of the main unit cell, which can be written as 
\begin{equation}
\label{eq:H_fiber}
	{\mathcal H} = {\mathcal H}_{\mathrm{self}} +\sum_{k = 1}^M {\mathcal H}_{\mathrm{int}}^{0k},
\end{equation}
where ${\mathcal H}_{\mathrm{self}}$ gives the {\em self-energy} of the main unit cell,  
and ${\mathcal H}_{\mathrm{int}}^{0k}$ accounts for the interactions between the main and
the $k$-th unit cells. 
The following interactions are included in these two terms:  
1) electrostatic interactions between all charged
beads along the PE chain with one another {\em and} with charged spheres within the linear Debye-H\"uckel
theory (which effectively captures salt screening effects) as well as the mutual interactions between the charged spheres themselves, 2) a short-ranged soft-core excluded-volume 
repulsion between chain beads and spheres only, which prevents the chain from penetrating the spheres, and
3)  the mechanical bending elasticity of the PE chain. Note that the electrostatic contribution to the bending
rigidity is explicitly included via the electrostatic self-energy term.
In the second term in Eq. (\ref{eq:H_fiber}), we have used a cut-off $M$ to 
account for the interaction between 
$M$ consecutive unit cells. We typically choose $M=5$. As explicitly checked, the results  
are not influenced by increasing $M$ further due to the highly screened nature of electrostatic 
interactions at high salt and/or highly expanded size of unit cells at low salt. 

\begin{figure*}[t]
\begin{eqnarray}
\label{eq:H_unitcell}  
{\mathcal H}_{\mathrm{self}}& =& \frac{\ell_{\mathrm{p}}}{\Delta}\sum_{i = 2}^N \bigg\{1-\cos(\theta_i-\theta_{i-1})
                           + \sin\theta_i\sin\theta_{i-1}\bigg(1-\cos(\phi_i-\phi_{i-1})\bigg)\bigg\}\\
                       & &+\,\,q^2\ell_{\mathrm{B}}\sum_{i=0}^{N-1}\sum_{j=i+1}^{N}\frac{e^{-\kappa |{\mathbf r}^0_i-{\mathbf r}^0_j|}}
                          {|{\mathbf r}^0_i-{\mathbf r}^0_j|}
		-\frac{Zq\ell_{\mathrm{B}}}{1+\kappa R_{\mathrm{s}}}\sum_{i=0}^{N}
			\bigg[\frac{e^{-\kappa( |{\mathbf r}^0_i|-R_{\mathrm{s}})}}{|{\mathbf r}^0_i|}-
				Ae^{-( |{\mathbf r}^0_i|-R_{\mathrm{s}})/\alpha}\bigg] \nonumber
\end{eqnarray}
\hrule
\end{figure*}
\begin{figure*}[t]
\begin{eqnarray}
\label{eq:H_interunit}
{\mathcal H}_{\mathrm{int}}^{0k} &=& q^2\ell_{\mathrm{B}}\sum_{i=0}^{N}\sum_{j=2}^{N}
	\frac{e^{-\kappa |{\mathbf r}^0_i-{\mathbf r}^k_j|}}{|{\mathbf r}^0_i-{\mathbf r}^k_j|}
			-\frac{Zq\ell_{\mathrm{B}} }{1+\kappa R_{\mathrm{s}}}
	\bigg[ \sum_{i=0}^{N} \bigg(\frac{e^{-\kappa (|{\mathbf r}^0_i-{\mathbf R}^k|-R_{\mathrm{s}}) }}{|{\mathbf r}^0_i-{\mathbf R}^k|}
	-A\,e^{-(|{\mathbf r}^0_i-{\mathbf R}^k| - R_{\mathrm{s}})/\alpha}\bigg)\nonumber\\
		&&+ \sum_{i=2}^{N} \, \bigg( \frac{e^{-\kappa ( |{\mathbf r}^k_i| - R_{\mathrm{s}}) }}{|{\mathbf r}^k_i|} 
				-A\,e^{-(|{\mathbf r}^k_i|-R_{\mathrm{s}})/\alpha} \bigg)\bigg] + 
	\frac{Z^2\ell_{\mathrm{B}}e^{2\kappa R_{\mathrm{s}}}}{(1+\kappa R_{\mathrm{s}})^2}\frac{e^{-\kappa|{\mathbf R}^k|}}{|{\mathbf R}^k|}, 
\end{eqnarray}
\hrule
\end{figure*}

The self-energy of the main unit cell is written in discretized form as
\begin{center}
{\em See equation (\ref{eq:H_unitcell}) above}
\end{center}
(in units of $k_{\mathrm{B}}T$), 
where the first term is the mechanical bending contribution, the second term is the 
electrostatic repulsion between chain  segments, and the third term represents chain-sphere attraction as
well as the soft-core repulsion (which is specified by a repulsion range, $\alpha$, and 
strength $A$, which are fixed at typical values of $A=0.014$~nm$^{-1}$ and 
$\alpha=0.02$~nm \cite{thesis}). 
Recall that  $\kappa=(8\pi \ell_{\mathrm{B}} C_s)^{1/2}$ is the inverse Debye screening length
 for monovalent salt with  $\ell_{\mathrm{B}} = e^2/(4\pi \varepsilon \varepsilon_0 k_{\mathrm{B}}T)$
 being the Bjerrum length. 

The interaction between the main unit cell and the $k$-th image unit cell reads (in units of $k_{\mathrm{B}}T$) 
\begin{center}
{\em See equation (\ref{eq:H_interunit}) above}
\end{center}
noting that the first two beads $i=0, 1$
for the $k$-th image unit cell are counted as the last beads of the $(k-1)$-th cell as explained before.  
The first term in Eq. (\ref{eq:H_interunit}) corresponds to repulsion between chain segments in the 
main unit cell and those in the $k$-th image cell, the second term involves electrostatic attraction and
soft-repulsion between the chain segment in the main unit cell and the sphere in  the $k$-th image cell and vice versa. 
The last term represents electrostatic repulsion between the spheres in the two unit cells.

\subsection{Numerical method: Unconstrained (full) minimization model}
\label{subsec:chromatin_fiber_min_method}

For strongly coupled PE-macroion complexes (i.e., with large polymer adsorption energy),  the 
so-called ground-state-dominance approximation becomes accurate \cite{Kunze2}.  
Here, one seeks the {\em optimal} or {\em ground-state} configuration of the fiber 
minimizing  the effective Hamiltonian (\ref{eq:H_fiber}) with respect to all degrees of freedom
that are necessary to specify the overall chain conformation. 
 This amounts to $2N+1$ independent degrees of freedom (per unit cell)
 $\{\theta_i, \phi_{i\neq P}, r_P, \chi\}$, where  $\theta_i$ and $\phi_{i\neq P}$ are the chain's bond vector 
 angles, $r_P$ is the $z$ component 
of the position of the middle bead of the chain $P=N/2$ (measured from the sphere center) and
$\chi$ the unit cell rotation angle. Note that  
$\phi_P$ (azimuthal angle of the subunit $P$ or bond vector ${\mathbf u}_P$) is fixed in order to 
remove Goldestone (zero) modes associated with a trivial rotational symmetry around the $z$ axis 
connecting the sphere center to the middle bead $P$. 

Due to a large number of free variables, we tackle the problem numerically using 
 the quasi-Newton optimization algorithm  \cite{Numerical}, which locates the minimum of the Hamiltonian using its first-order derivatives.
 At a minimum, the first-order variation of the Hamiltonian vanishes, i.e. 
\begin{equation}
\delta {\mathcal H}  =
\sum_{i = 1}^N\bigg(\frac{{\mathrm{d}} {\mathcal H} }{{\mathrm{d}}\theta_i} \delta\theta_i+ \frac{{\mathrm{d}} {\mathcal H} }{{\mathrm{d}}\phi_{i\neq P}}\delta\phi_{i\neq P}\bigg) + 
\frac{{\mathrm{d}} {\mathcal H} }{{\mathrm{d}}\chi} \delta\chi + \frac{{\mathrm{d}} {\mathcal H} }{{\mathrm{d}}r_P}\delta r_P = 0.
\end{equation}
                     
We use a combination of stochastic and parameter quenching methods to make sure that 
the obtained structures are not meta-stable (local minima). The meta-stable 
structures have been discussed elsewhere for chain-sphere complexes \cite{Kunze2,thesis}.

\begin{figure*}[t]
\begin{center}
\includegraphics[angle=0,width=17.cm]{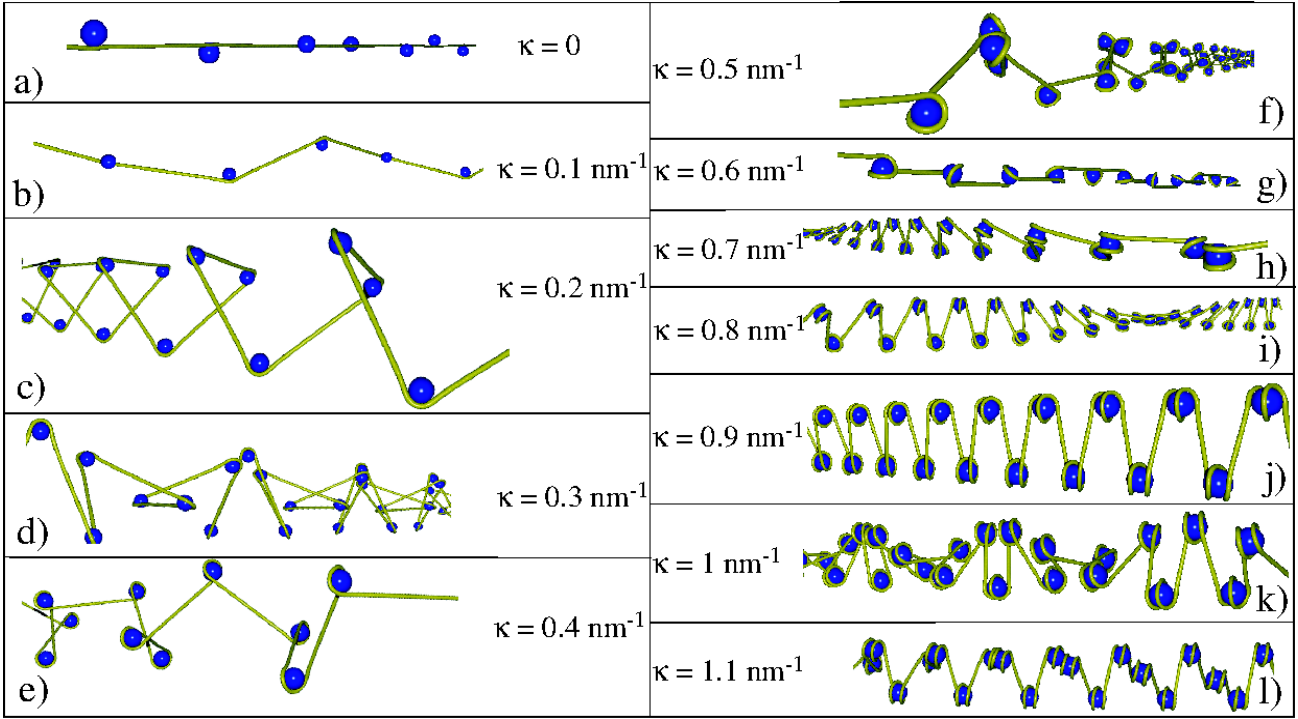}
\smallskip
\caption{\label{fig:salt_var}The optimal spatial configuration of the complex fiber 
as calculated from the unconstrained minimization of the Hamiltonian Eq.  (\ref{eq:H_fiber})
for macroion (sphere) charge valency $Z=15$, total chain length per unit cell (per sphere) of $L_c=68$~nm 
(equivalent to 200 DNA base pairs) and for various Debye inverse screening lengths as indicated
on the graph.}
\end{center}
\end{figure*}


\section{ Salt-induced structural changes of the fiber - unconstrained optimization}
\label{sec:fiber_salt}

In this section, we investigate the dependence of structural properties of the complex fiber
on the salt concentration for fixed macroion charge valency $Z=15$ and 
fixed  total length of chain per unit cell $L_c=68$~nm (equivalently 
$N_{\mathrm{bp}}=200$ base pairs of DNA) \cite{Thoma,Bednar}. 
The salt concentration $C_s$  is increased from zero up to the physiologically relevant range of 
100 mM monovalent salt and beyond (with the inverse screening length spanning the range 
$\kappa=0-1.1$~nm$^{-1}$). 
Here we focus on the unconstrained (full) minimization scheme; 
therefore, the chain length wrapped around the macroion  is not fixed and can change accordingly. 
The  results display  a variety of helical fibers as visualized in Fig. \ref{fig:salt_var}, 
which  reflect drastic structural variations by changing the salt concentration. 

The ground-state configurations shown in the Figure can be grouped into a few generic classes, 
such as  {\em beads-on-a-string} (Figs. \ref{fig:salt_var}a, b, g), 
{\em zig-zag} or cross-linker (Figs. \ref{fig:salt_var}d, h-l) and {\em loose solenoidal structures } 
(Figs. \ref{fig:salt_var}c, e, f). These types of patterns  have been identified more specifically within recent geometric models for the 
chromatin fiber \cite{Schiessel_rev,Woodcock,Wedemann}. In general, the zig-zag pattern can be thought of as a configuration in which successive 
core particles are located on opposite sides of the fiber and the straight linker chain  joining them bridges 
across the helical core of the fiber. This property is absent in the  loose solenoidal pattern where successive nucleosomes are adjacent (as nearest neighbors). 
Note that this classification is only qualitative in the present context and will be used for convenience when referring to the 
structures obtained from our model. A more quantitative analysis of the salt-dependent behavior will be discussed in the following Sections.   

\begin{figure*}[t]
\begin{center}
\includegraphics[angle=0,width=15.cm]{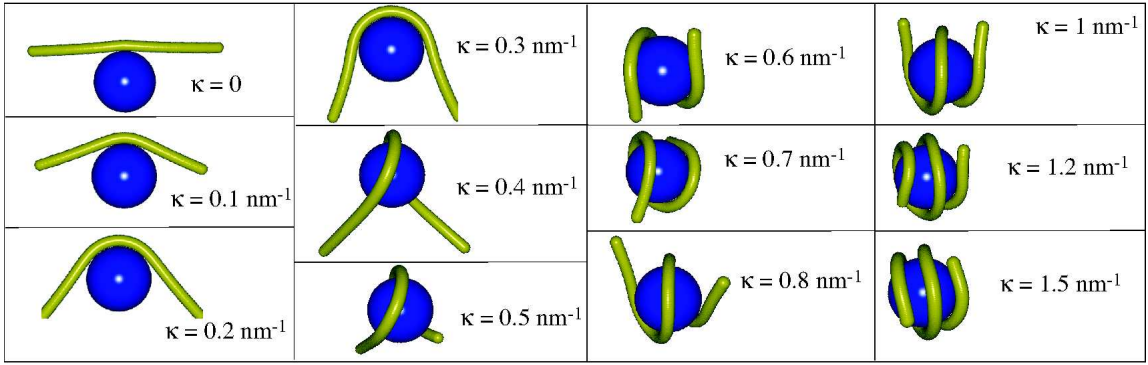}
\smallskip
\caption{\label{fig:unitcell}The local shape of the core particles within the optimal fiber structures in Fig. \ref{fig:salt_var} 
for sphere charge valency $Z = 15$ and various Debye inverse screening lengths as indicated 
on the graph. The total length of the PE chain per unit cell is fixed as $L_c = 68$~nm ($N_{\mathrm {bp}} = 200$). 
The increase in the number of chain
turns around the macroion is clearly demonstrated. }
\end{center}
\end{figure*}

\begin{figure}[t]
\begin{center}
\includegraphics[angle=0,width=8.cm]{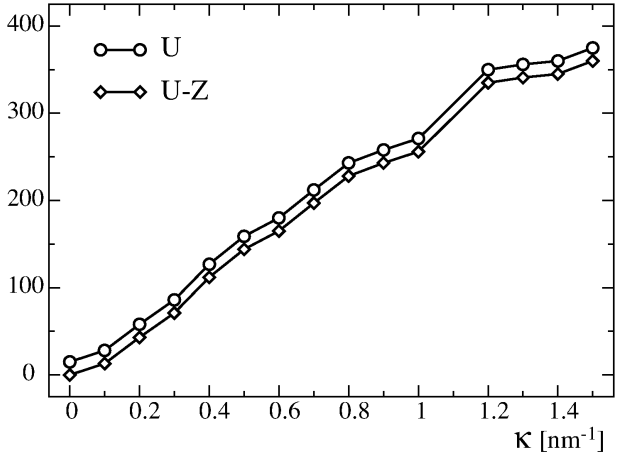}
\smallskip
\caption{\label{fig:overcharge}Number of chain beads (circles) that are attached to the sphere 
(i.e., within the distance 1.02$R_{\mathrm{s}}$ from the sphere center where $R_{\mathrm{s}}=5$~nm is the sphere radius)
as a function of the inverse screening length for the structures shown in Figs. \ref{fig:salt_var} and \ref{fig:unitcell}. 
As noted before each bead is monovalent ($q=1$), thus this number equals the PE charge adsorbed on the sphere. 
Diamonds show the net charge of the complex. At very low salt concentration, the sphere charge is 
already nearly compensated by the adsorbed PE segment and for larger salt, 
the sphere is strongly overcharged.}
\end{center}
\end{figure}

As seen, the degree of chain complexation increases with increasing salt concentration. 
The highly wrapped states occur at intermediate 
salt concentration and result from the growing dominance of electrostatic chain-sphere 
attraction against the chain (electrostatic and bending) self-energy. Indeed, as we 
show later,  macroions
are highly {\em overcharged} by the PE adsorption much in the same way as a single isolated 
complex is overcharged at intermediate to large salt concentration \cite{Kunze2,hoda1,Park,Shklovs02}. 

At vanishing salt concentration (Fig.  \ref{fig:salt_var}a), the PE chain takes a straight-line shape due to 
the dominant electrostatic self-energy of the PE chain as charged units interact with 
long-ranged (unscreened) Coulomb interaction. 
The distribution of spheres around the chain
is periodic  but their centers lie on different angular locations around the chain. 
For inverse screening length in the range of $\kappa=0.3$ to 0.6~nm$^{-1}$ (Figs. \ref{fig:salt_var}d-g), the 
fiber structure shows an anomalous behavior, i.e., the diameter of the fiber decreases and the projected distance 
(along the fiber axis) between neighboring sphere centers increases, which is due to the fact that the chain strand
completes its first turn around each sphere (see also Fig. \ref{fig:unitcell}). 
Upon further complexation of the chain (Figs. \ref{fig:salt_var}h-l), the fiber thickness tends to 
increase again and the spheres become more densely packed. 
Note that in the range of Debye inverse  screening lengths $\kappa= 0 - 0.5$~nm$^{-1}$
($<$25 mM monovalent salt), the chain wrapping is less than a complete turn and these
configurations appear locally as beads-on-a-string structures. 
However,  the chain wrapping may be increased by taking a higher sphere charge, leading
to different structures  even for small $\kappa$,  as shown elsewhere \cite{thesis}. 
The role of sphere charge will be discussed later. 

\subsection{Overcharging of the core particle}

It is useful to consider the local structure of the core particles within the fiber in more detail. 
A closer view of the unit cell of the fiber is shown in Fig. \ref{fig:unitcell} for different salt concentrations. 
We introduce two quantities that measure the degree of PE chain wrapping around each macroion in the fiber, 
namely, the {\em adsorbed} PE  charge, $U$ (that includes only those chain beads that are within 
distances less than or equal to $1.02R_{\mathrm{s}}$ from the sphere center), 
and the net charge of the core particle defined as $U-Z$. Both quantities  are shown in Fig. \ref{fig:overcharge}
as a function of $\kappa$.

The unit cell configurations clearly show that the PE chain becomes gradually more wrapped around the sphere
upon increasing the salt concentration. That amounts to more than two complete turns at high salt ($\kappa=1.5$~nm$^{-1}$). 
As seen from Fig. \ref{fig:overcharge}, the adsorbed PE charge increases almost linearly with the salt concentration 
and  reaches a plateau-like region at high salt in agreement with the behavior observed for a single
isolated complex \cite{thesis}. 
The saturation at high salt  is due to the finite and fixed  length of the PE chain per sphere that is 
$200$ DNA base pairs in the present case. The adsorbed PE charge amounts to that of a DNA segment of 146 base pairs
at about the physiological salt concentration  ($\kappa\simeq 1$~nm$^{-1}$) \cite{Kunze2}. 
  
 Note that for the chosen parameters here
(with sphere valency of $Z=15$), the core particle is always overcharged, that is $U-Z>0$. 
The overcharging degree, $(U-Z)/Z$,  
becomes quite large at intermediate salt, such that the net complex charge becomes  
more than twenty times larger in magnitude than the bare sphere charge.
This striking feature is known to occur due to strong lateral chain segment 
correlations at large salt concentration, which leads to a highly ordered adsorbed layer
on the macroion surface \cite{Kunze2,hoda1,Park,Shklovs02,Netz-JF}.
In this situation, the spacing between chain strands on the sphere is of the order of 
the Debye screening length, $\kappa^{-1}$ \cite{Netz-JF}. 

The unit cell configurations in Fig. \ref{fig:unitcell} also display how the relative entry-exit angle of the chain in a core particle  
is influenced by the changes in the salt concentration. The changes in the relative entry-exit angle  in turn affects the overall arrangement of the 
core particles and thus the global structure of the fiber as seen in Fig. \ref{fig:salt_var}.

\begin{figure*}[t]
\begin{center}
\includegraphics[angle=0,width=11.cm]{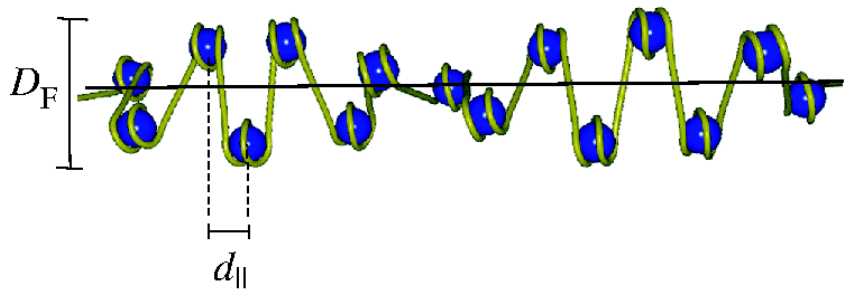}
\smallskip
\caption{\label{fig:fiber_analys}Schematic representation of the fiber diameter,  $D_{\mathrm{F}}$, and 
the projected center-to-center distance, $d_{||}$,  of consecutive spheres along the fiber axis. 
The central fiber axis is shown by a horizontal solid line.}
\end{center}
\end{figure*}

\begin{figure*}[t]
\begin{center}
\includegraphics[angle=0,width=14.cm]{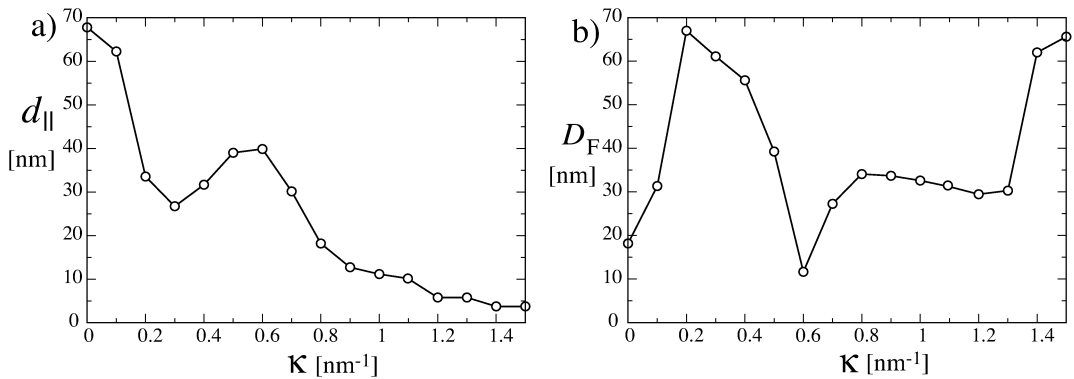}
\smallskip
\caption{\label{fig:diacon_kappa}a) The projected distance of neighboring sphere centers along the fiber axis, $d_{||}$, 
and b) the fiber diameter, $D_{\mathrm{F}}$, as a function of the inverse screening length. 
The macroion charge and chain length per unit cell are fixed as $Z = 15$ and $L_c = 68$~nm.}
\end{center}
\end{figure*}

\subsection{Geometry of the fiber}
\label{subsec:fiber_salt_sum}

\begin{figure*}[t]
\begin{center}
\includegraphics[angle=0,width=9.cm]{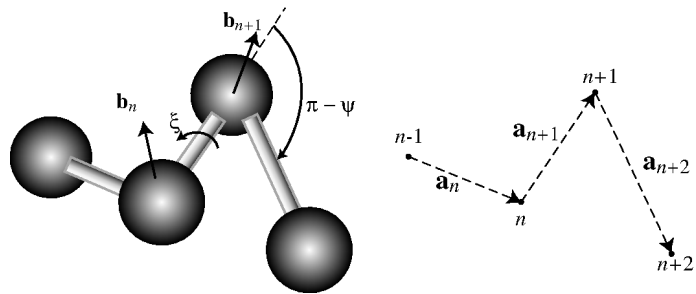}
\smallskip
\caption{\label{fig:wood_model}A schematic view of the two-angle representation of macroion positions along the chromatin fiber. 
Each two successive spherical macroions, labeled by $n-1$ and $n$, are connected to 
each other via a vector ${\mathbf a}_n$. The entry-exit angle 
$\psi$ is defined as the complementary angle between any two consecutive connecting vectors (shown in the figure for ${\mathbf a}_{n+1}$ and ${\mathbf a}_{n+2}$),
which is the same for all $n$ in a homogeneous helical structure as is the case in the present model. On the other hand, each three successive macroions (e.g., 
those labeled by $n-1$, $n$ and $n+1$) define a plane with a normal vector ${\mathbf b}_n= {\mathbf a}_n\times{\mathbf a}_{n+1}$.
 The dihedral angle $\xi$ is defined as the 
angle between two such normal vectors (shown in the figure for ${\mathbf b}_n$ and ${\mathbf b}_{n+1}$).}
\end{center}
\end{figure*}

Besides the overall shape,  one is interested in geometrical characteristics 
of the fiber and the way they vary with the salt concentration. 
Here we consider the following parameters (see Fig. \ref{fig:fiber_analys}):
\begin{itemize}
\item[-] The fiber diameter,  $D_{\mathrm{F}}$, which refers to the diameter of the smallest {\em outer} 
cylinder  coaxially enclosing the fiber. 
\item[-] The projected center-to-center distance, $d_{||}$, of consecutive spheres along the central fiber axis, which
gives a measure of {\em projected density} of the fiber, $n_{||}$; conventionally, we represent the fiber density 
as the number of spheres per 11~nm as typically expressed in experimental literature for chromatin \cite{Bednar}; i.e. 
\begin{equation}
n_{||}=\frac{11}{d_{||}},   
\end{equation} 
where $d_{||}$ is measured in units of nm. 
\item[-] The entry-exit angle, $\psi$, and the dihedral angle, $\xi$, which can be defined  (Section \ref{sec:chromatin_constrained}) 
in analogy with the chromatin models  \cite{Woodcock} (see Fig. \ref{fig:wood_model}). 
\end{itemize}


As seen in Fig. \ref{fig:diacon_kappa}a, the projected distance takes a maximum value of about $d_{||}=68$~nm
at  zero salt concentration and then falls off in a non-monotonic fashion to quite small values of about sphere 
radius $d_{||} \simeq R_{\mathrm{s}}=5$~nm, when the inverse screening length 
(salt concentration)  exceeds 
$\kappa=1$~nm$^{-1}$ (100 mM NaCl) representing  a dense fiber of a density about $n_{||}=2$. 
The projected distance  displays
a local minimum and a local maximum at inverse screening lengths of about $\kappa=0.3$~nm$^{-1}$ 
and $\kappa=0.6$~nm$^{-1}$, respectively. These extrema roughly correspond to extrema  
in the fiber diameter,  $D_{\mathrm{F}}$, which shows an almost opposite trend as compared with 
the projected distance when the salt concentration is varied (Fig. \ref{fig:diacon_kappa}b). 
At zero salt, the outer fiber diameter is roughly twice the sphere diameter, i.e., $D_{\mathrm{F}}=20$~nm, due to a staggered configuration of the spheres. 
It changes rapidly at small salt concentration with $\kappa$, reaching a maximum and then 
a sharp minimum at an intermediate salt concentration of about   
$\kappa=0.6$~nm$^{-1}$, where the projected distance exhibits a local maximum.
The minimum fiber diameter at this point is about 10~nm (equal to the sphere diameter); hence at
$\kappa=0.6$~nm$^{-1}$, the complex fiber shows its largest projected distance to diameter ratio
$d_{||}/D_{\mathrm{F}} = 4$, reflecting an almost one-dimensional 
beads-on-a-string structure  (see Fig. \ref{fig:salt_var}g).

Interestingly, over the range of intermediate salt concentrations between $\kappa=0.8$ and 1.3~nm$^{-1}$ (corresponding
to 60-160 mM monovalent salt), the fiber diameter remains almost unchanged at about 
$D_{\mathrm{F}}=30$~nm (reflected by a plateau-like region in Fig. \ref{fig:diacon_kappa}b), 
although the fiber structure changes from the zig-zag to loose solenoidal pattern (see also Fig. \ref{fig:chromatin_phase_kappa}). 
The chain wrapping in this region
amounts to about 1-and-3/4 turn, close to that in the nucleosome core particles in
chromatin at intermediate salt concentration about the physiological regime \cite{The_cell,Thoma,Woodcock,Bednar}. 
In fact, the experiments reveal that the chromatin
fiber exhibits a dense structure in this regime with a diameter of about 30~nm, the so-called  ``30~nm fiber". 
The present results, which predict a stable 30~nm complex fiber at intermediate salt concentration, thus appear to give 
a trend consistent with those found in chromatin experiments \cite{The_cell,Thoma,Woodcock,Bednar}. The present full-minimization model
however is very different from the H1-stabilized chromatin because (aside from other realistic factors in chromatin that are missing in our model) 
the structure of the core particle (and the conformation of the entering and exiting chain strands) can change subject to interactions with other units.
Thus the local fiber structure 
is different from that of chromatin. 
We also obtain a  lower density of spheres ($n_{||}\leq 1$) as compared with the chromatin fiber. 
The present results however suggest that such 30~nm fibers may be feasible in mixtures containing DNA and synthetic nano-colloids 
at  intermediate salt concentrations. 

For large salt concentrations (beyond $\kappa=1.3$~nm$^{-1}$), the fiber diameter rapidly increases 
but the projected distance between spheres remains constant. At $\kappa=1.5$~nm$^{-1}$, the fiber exhibits 
a small ratio of $d_{||}/D_{\mathrm{F}}=0.077$, corresponding to a tight solenoidal shape 
(shown on Fig. \ref{fig:chromatin_phase_kappa}).

\subsection{Two-angle structural  diagram}

The arrangement of {\em macroions} along the fiber within the present model 
may also be described using a two-angle description in the spirit of the  
two-angle model presented in Ref. \cite{Woodcock} for the chromatin fiber.   
For the sake of analogy, we define an  entry-exit angle,  $\psi$,  and a dihedral angle,  $\xi$, 
as follows  (see Fig. \ref{fig:wood_model}). 
 
Let us assume that ${\mathbf a}_n = {\mathbf R}^{n} - {\mathbf R}^{n-1}$ is the connecting vector between 
two successive macroion centers, where ${\mathbf R}^n$ is the position of the $n$-th sphere. 
The entry-exit angle, $\psi$, is defined from  the angle between two such consecutive 
connecting vectors, ${\mathbf a}_n$ and ${\mathbf a}_{n+1}$, i.e.
\begin{equation}
\label{eq:psi}
\psi = \pi -\cos^{-1} \bigg(\frac{{\mathbf a}_n\cdot{\mathbf a}_{n+1}}{|{\mathbf a}_n||{\mathbf a}_{n+1}|}\bigg).
\end{equation}

\begin{figure*}[t]
\begin{center}
\includegraphics[angle=0,width=14.cm]{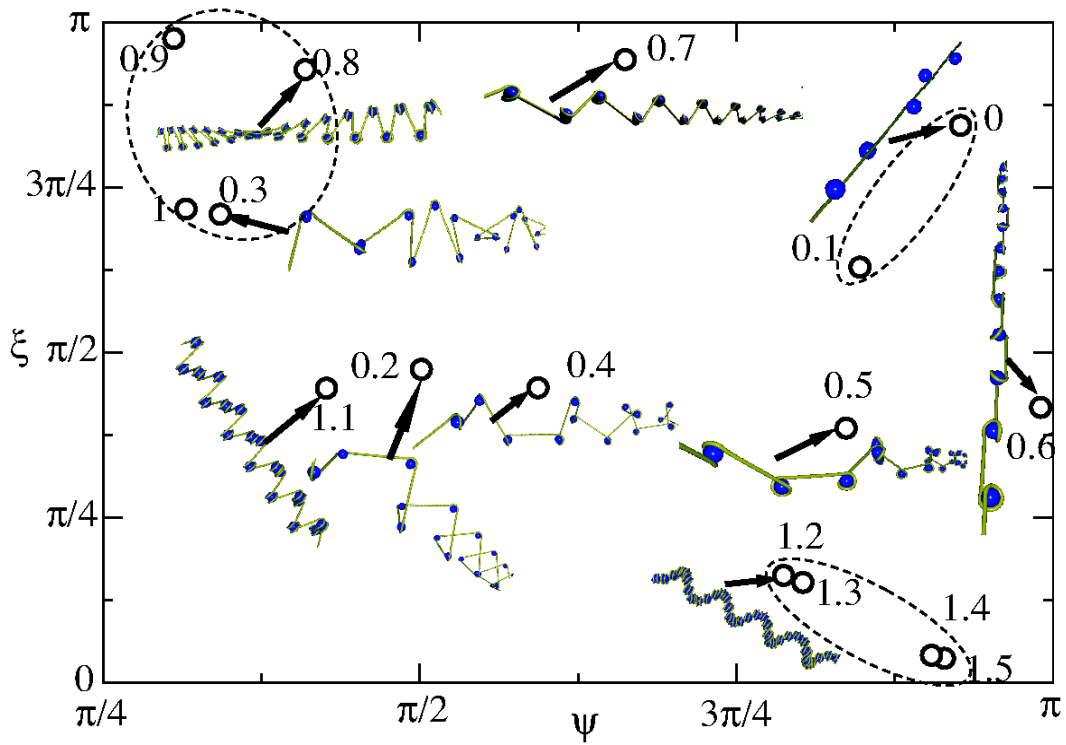}
\smallskip
\caption{\label{fig:chromatin_phase_kappa}The two-angle diagram for fiber structures from
unconstrained optimization at fixed macroion charge valency 
$Z = 15$, fixed chain length per unit cell $L_c = 68$~nm ($N_{\mathrm {bp}} = 200$), and for various $\kappa$.  
The value of $\kappa$ associated with each point is indicated on the graph (in units of nm$^{-1}$). The zig-zag structures
(upper left corner), beads-on-a-string structures (upper right region),
and compact solenoids (lower right region) are indicated schematically by closed ellipses and 
for each case a characteristic  structure is shown. }
\end{center}
\end{figure*}

The dihedral angle, $\xi$, the angle by which four consecutive spheres are out-plane,  is determined  
 by calculating the angle between the normal vectors of two planes, one   identified by the
spheres ${\mathbf R}^{n-1}$, ${\mathbf R}^{n}$, ${\mathbf R}^{n+1}$ and the other identified by ${\mathbf R}^{n}$, 
${\mathbf R}^{n+1}$, ${\mathbf R}^{n+2}$. Explicitly, one has 
\begin{equation}
\label{eq:torsion}
\xi = \cos^{-1} \bigg(\frac{{\mathbf b}_n\cdot{\mathbf b}_{n+1}}{|{\mathbf b}_n||{\mathbf b}_{n+1}|}\bigg)
\end{equation}
with ${\mathbf b}_n = {\mathbf a}_n\times{\mathbf a}_{n+1}$.
In a homogeneous helical structure, which is the case in the present model,  $\psi$ and $\xi$ are 
independent of $n$ and thus the same angles are obtained between any two consecutive vectors.

\begin{figure*}[!t]
\begin{center}
\includegraphics[angle=0,width=14.cm]{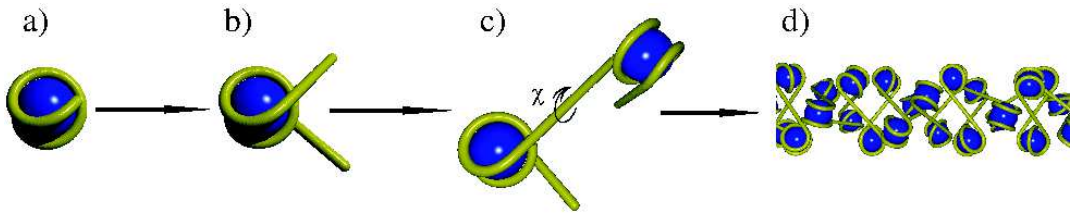}
\smallskip
\caption{\label{fig:mode_rotate}In the constrained optimization model, 
the conformation of the chain in each unit cell is fixed and only
the rotational angle $\chi$ can vary.   The core particle complex (a) consists of a chain segment of length equivalent 
to 146 DNA base pairs, whose configuration is obtained from numerical minimization of a single
isolated complex for a given sphere charge and salt concentration. 
The core particles are linked together with straight linker chain of arbitrary length (b) (in the shown configuration, the
total chain length per unit cell is $N_{\mathrm{bp}}=200$). The replication
procedure (c and d) is the same as explained in Section \ref{subsec:chromatin_periodic}.}
\end{center}
\end{figure*}

By calculating these two angles 
for various salt concentrations, one can sketch a two-parameter $\xi-\psi$ structural diagram. 
The result is shown in Fig. \ref{fig:chromatin_phase_kappa} for fixed sphere 
charge $Z = 15$ and fixed chain length (per unit cell) of  $N_{\mathrm{bp}} = 200$. 
Open circles give the two-angle coordinates associated with the numerically calculated optimal 
configuration at given inverse screening lengths (indicated by numbers on the graph). 

It follows from the diagram that the angle $\psi$ determines the compactness of the 
fiber, while $\xi$ roughly determines its structural class, e.g., being zig-zag or solenoidal. 
At small salt concentrations, these angles vary rapidly with the salt concentration, 
which, as mentioned above, is  a consequence of the chain wrapping process. At high salt, most
of the free length of the chain available in a unit cell is already wrapped and thus the angles vary
weakly.  The parameter which varies more dramatically with $\kappa$ is $\psi$. 
As seen, $\psi$ never becomes smaller than $\pi/4$, due to excluded-volume interactions 
\cite{Schiessel_rev}, while $\xi$ spans the whole interval  $0< \xi < \pi$. 

The beads-on-a-string patterns are found at the boundaries of the diagram with entry-exit
angles roughly in the range  $3\pi/4 <\psi< \pi$. While the compact solenoidal structures appear in the lower right corner, 
that is with small dihedral angle $0<\xi < \pi/4$ but large entry-exit angle $3\pi/4<\psi < \pi$.  
One may thus distinguish {\em compact solenoidal} structures on the lower-right  corner of the graph.  
The 30~nm zig-zag structures obtained at intermediate salt regime typically appear on the left upper corner of the
graph, that is for $\pi/4 < \psi <\pi/2$ and typically for large $\xi>\pi/2$.

\begin{figure*}[!t]
\begin{center}
\includegraphics[angle=0,width=13.cm]{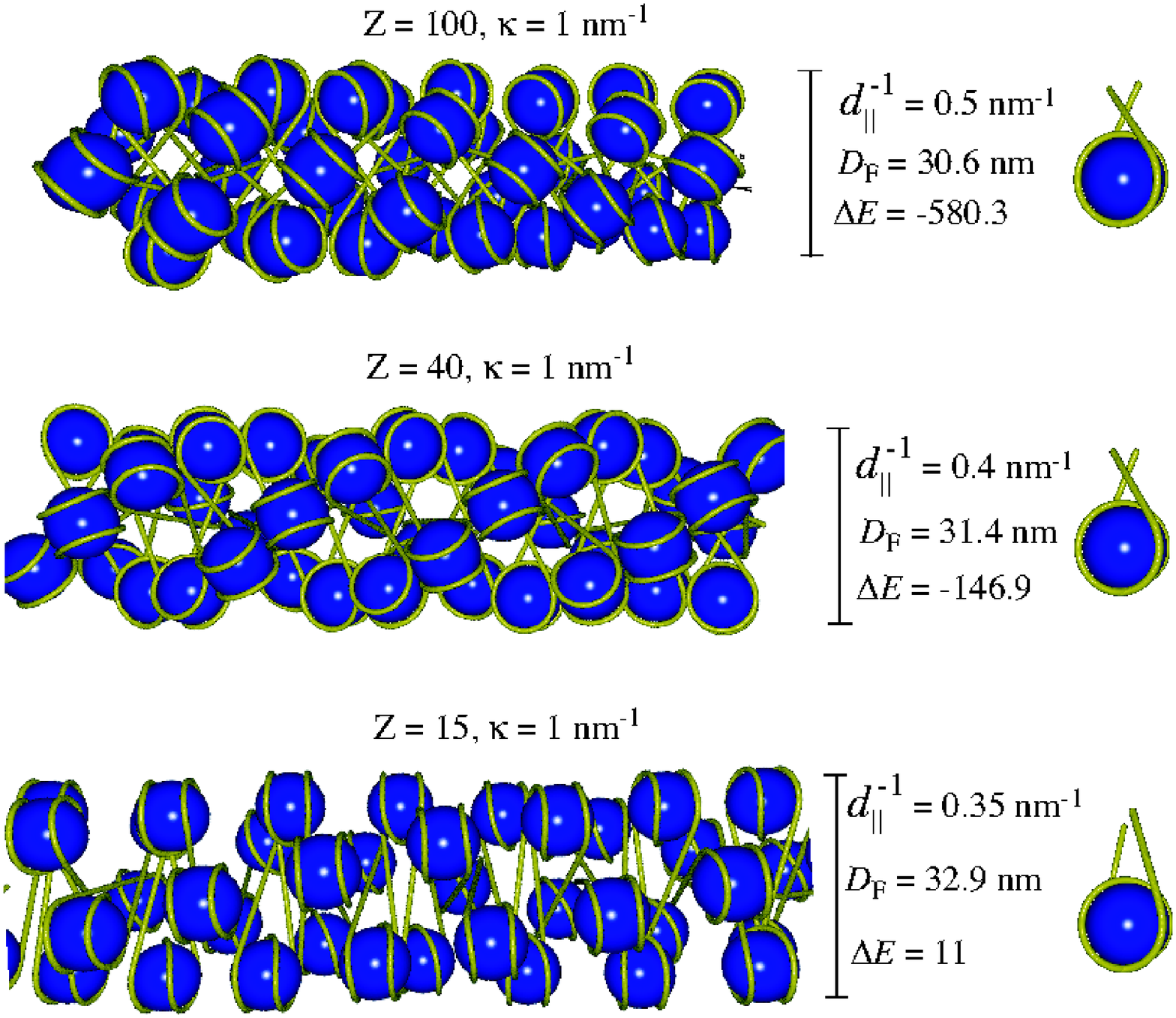}
\smallskip
\caption{\label{fig:chromatin_rotation_Zgal}Optimal fiber configurations obtained from the constrained optimization model for 
different  sphere charge $Z$ as indicated on the graph but for 
fixed  inverse screening length $\kappa = 1$~nm$^{-1}$ and PE chain length per unit cell $N_{\mathrm{bp}}=200$.
The inverse projected distance, $d_{||}^{-1}$, 
the binding energy, $\Delta E$ (in units of $k_{\mathrm{B}}T$), and the fiber diameter $D_{\mathrm{F}}$ are given in the graph for each fiber.
The corresponding local structure of the fiber unit cell is shown on the right.}
\end{center}
\end{figure*}

\begin{figure*}[!t]
\begin{center}
\includegraphics[angle=0,width=13.cm]{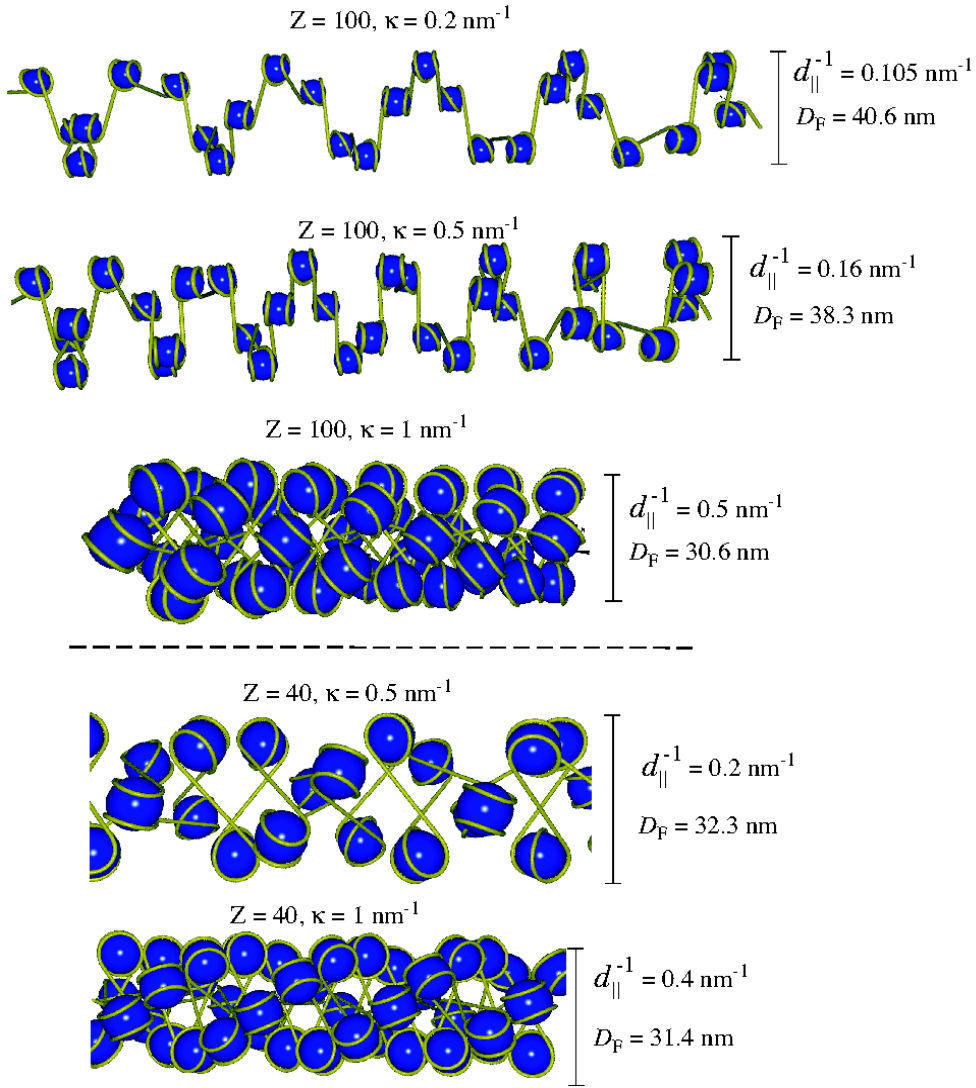}
\smallskip
\caption{\label{fig:rotation_2} Same as Fig. \ref{fig:chromatin_rotation_Zgal} but for fixed sphere charge $Z=100$ and 40 and different  
 inverse screening lengths as indicated on the graph. }
\end{center}
\end{figure*}

\begin{figure*}[ht]
\begin{center}
\includegraphics[angle=0,width=12.cm]{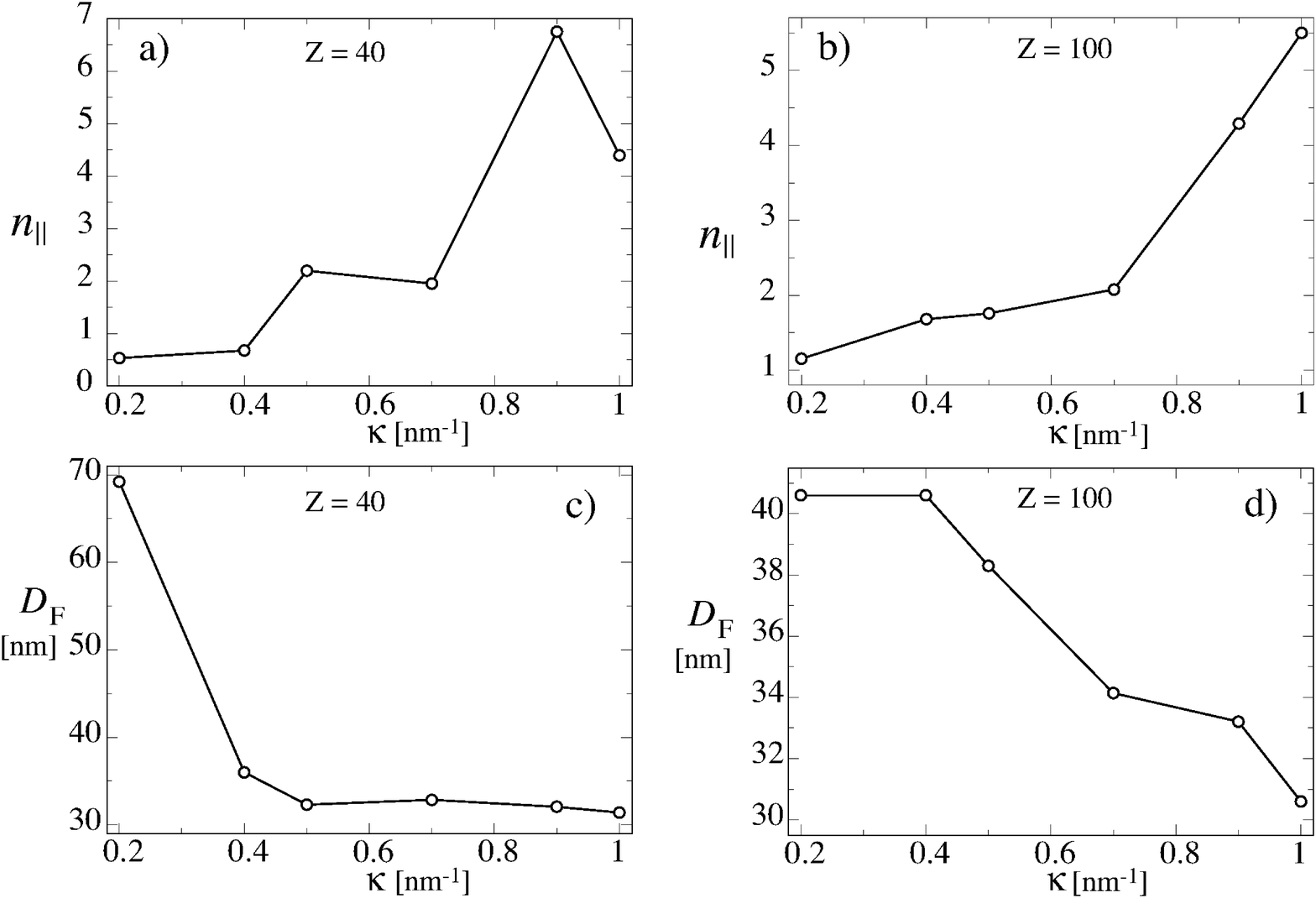}
\smallskip
\caption{\label{fig:rotation_Dnvck}a) and b) Density of spheres (number of spheres per $11$nm along the fiber axis) $n_{||}$, 
and c) and d) the fiber diameter, $D_{\mathrm{F}}$, as a function of the inverse screening length as obtained from the constrained optimization model. 
The macroion charge is $Z = 40$ for a) and c) and $Z = 100$ for b) and d). The chain length per unit cell is fixed at 200 DNA 
base pairs ($L_c = 68$~nm).}
\end{center}
\end{figure*}


\section{ Constrained optimization model}
\label{sec:chromatin_constrained}


As noted before, several features of the actual chromatin structure are not accounted for within the present model. 
One  factor is the so-called {\em linker histone}
H1, a cationic protein which binds near the entry-exit region of the DNA in the nucleosome. 
It brings together the two strands of the DNA that enter and exit the central core particle 
forming a stem-like pattern \cite{Thoma,Bednar}. 
In  {\em in vitro} experiments, H1 histone can be removed by exposing the system
to  high salt concentrations (about 0.6 M), and washing the solution in order to remove dissociated linker 
histone. In such H1-depleted cases,
the chromatin fiber appears to be rather open and loose
(i.e., with a low number of histones per unit length along the fiber) \cite{Thoma}. 
While in the experiments with H1-stabilized linker histone, 
the core particle configuration is fixed and no unwrapping 
of the DNA on the histone
octamer is possible \cite{Bednar}. As a result, the number of base pairs complexed with the histone octamer is fixed.  

Motivated by this observation, we 
introduce a variant of the preceding model, in which the structure of the core
chain-sphere complex is constrained to be fixed. At any given value of the macroion charge $Z$ and the inverse screening length $\kappa$, we fix the core structure within each unit cell
according to the optimal (ground-state) configuration  of an {\em isolated} complex (with a total chain segment of length about 146 DNA base pairs).  
This structure is obtained from a similar minimization study as discussed in Refs. \cite{Kunze2,hoda1}.
We then connect straight linker chains of arbitrary length to the
strands of the chain  entering and exiting  the core particle. This constitutes a single unit cell for the fiber
which is hence constructed by replicating this unit cell using the  transformation method
explained in Section \ref{subsec:chromatin_periodic}--see Fig. \ref{fig:mode_rotate}. 
Note, however, that in this case, there is only {\em one} degree of freedom, 
that is the angle $\chi$, which can be varied in order to find the minimum of the effective Hamiltonian 
of the fiber  (see Sections \ref{subsec:chromatin_fiber_interaction} 
and  \ref{subsec:chromatin_fiber_min_method}). 
Note also that the entry-exit angle is dictated by the configuration of a single complex in {\em isolation},
which thus depends only on  the sphere charge, $Z$, and the salt concentration via $\kappa$,
but not on the linker length.

\subsection{Optimal constrained configurations for various sphere charge}
\label{subsec:chromatin_varZ}

In Fig. \ref{fig:chromatin_rotation_Zgal},  we show several  structures obtained from the constrained minimization 
for the inverse screening length $\kappa = 1$~nm$^{-1}$ and for large to small  sphere charge $Z$ 
(at fixed chain length $N_{\mathrm{bp}}=200$). 
It is seen that the compactness of the fiber is less sensitive to changes in the sphere charge. 
As a result,  the entry-exit angle of the PE strand changes weakly as compared with the unconstrained minimization case.
The fiber becomes denser for large sphere charge: For $Z=15$, we have $d_{||}^{-1}=0.35$~nm$^{-1}$
corresponding to $n_{||}=3.85$ spheres
per 11~nm length along the fiber axis, while for
$Z=100$, we have  $n_{||}=5.5$  ($d_{||}^{-1}=0.5$~nm$^{-1}$), 
very close to the experimental chromatin value of 
 about 5-6 histones per 11~nm as reported in Refs. \cite{Thoma,Gerchman,Woodcock,Bednar,Horowitz}. 

In order to understand the underlying mechanism behind the compaction of the fiber with increasing the sphere charge, 
we consider the binding energy of the fiber, $\Delta E$, i.e., the energy of the optimal configuration measured with respect
to the reference state of a free straight PE chain in the absence of attached spheres. 
While for $Z=15$, the binding energy is positive (which reflects the fact that we are performing a constrained optimization
which does not necessarily find the global energy minimum), it becomes 
highly negative for larger $Z$ as indicated in Fig. \ref{fig:chromatin_rotation_Zgal}, i.e., $\Delta E=-580.3$ (in units of $k_{\mathrm{B}}T$) for $Z=100$. 
This in fact reflects effective attractive interactions between core particle complexes within the fiber, which result from
correlations between positive and negative patches on neighboring core complexes as may be seen
in the spatial configurations. Such inter-complex correlations have been studied before \cite{hoda1}. 

Note also that the diameter of the fiber is quite stable at about $D_{\mathrm{F}}=30$~nm, that is about the diameter
of the H1-intact 30~nm chromatin fiber   in the physiological salt regime (i.e., $\kappa\simeq 1$~nm$^{-1}$).
In fact, the striking point here is that within this simple model, the entering and exiting strands of the PE
chain form a cross pattern as explicitly shown by focusing on a 
single unit cell in Fig.  \ref{fig:chromatin_rotation_Zgal} (shown on the right). 
This situation--that resembles the experimentally observed stem-like pattern for the DNA in the nucleosome--is 
only obtained within the constrained minimization and leads to a small entry-exit angle, and hence a 
 rather compact fiber in qualitative agreement with the Woodcock model \cite{Woodcock}. 

\subsection{Optimal constrained  configurations for various salt concentration}
\label{subsec:chromatin_var_kappa}

In Fig. \ref{fig:rotation_2}, we show configurations at fixed $Z$ and different  salt concentrations. 
As seen for $Z=100$, the structure of the fiber becomes expanded for small $\kappa$ as a direct consequence of
 the increased electrostatic repulsion between the spheres. The density
of the fiber reduces to about one sphere per 11~nm and the fiber diameter increases to about 40~nm as indicated on the graph. 

As discussed elsewhere \cite{Kunze2,hoda1},
the sphere charge should be taken as an effective (or fitting) 
parameter within the present model, because the true sphere charge is regulated
under experimental
condition  depending
on the $p$H, or it may shift from the bare value due to charge renormalization effects in a salt solution \cite{note_tau_max}.
In fact, within the present model, the most relevant fiber configurations (that might resemble the chromatin fiber best)
are obtained by changing $Z$ according to the salt concentration. This may be thought of as a  fitting procedure,
since as explained above, the effective sphere charge is expected to depend on $\kappa$. To make this point clear, note that
for instance the fiber configurations for $Z=100$ and low $\kappa$ do not show zig-zag patterns
 \cite{Thoma,Allan,Bednar,Woodcock,Horowitz}.  
But zig-zag patterns at small $\kappa$ may be recovered by taking a smaller 
value for the sphere charge $Z$ (compare $\kappa=0.5$~nm$^{-1}$ for $Z=40$ and 100 in the Figure). 

To summarize these results, we plot the the fiber diameter, $D_{\mathrm{F}}$, and the fiber 
sphere density, $n_{||}$,  in  Fig. \ref{fig:rotation_Dnvck} for two different values of sphere charge $Z=40$ and 100.
Qualitatively, it appears that the sphere density increases from about one at low salt up to
5-6 spheres  in the physiological regime of  $\kappa\simeq1.0$~nm$^{-1}$ (100mM monovalent salt) \cite{Schiessel_rev,Thoma,Gerchman,Woodcock,Bednar,Horowitz}. 
 Examining the fiber diameter in  Figs. \ref{fig:rotation_Dnvck}c and d, 
we find that the fiber diameter decreases down to almost exactly 30~nm at the physiological regime. 
It is noteworthy that the arrangement of the spheres in the resultant 30~nm fiber shows a clear double helix pattern for $Z=40$ and $\kappa=1$~nm$^{-1}$.

\section{\label{sec:length_var} Variation of the chain length - unconstrained optimization}

In this Section, we return to the unconstrained optimization model and 
 address the effects due to changes in the length of the 
PE chain per unit cell, $N_{\mathrm{bp}} $,  on the fiber structure. 
We change $N_{\mathrm{bp}} $ in the range of 50 to 240 DNA base pairs 
and focus on the inverse Debye  screening length 
$\kappa = 0.8$~nm$^{-1}$ and the sphere charge valency $Z = 15$. 
At this salt concentration, the fiber exhibits a   30~nm zig-zag pattern (Fig. \ref{fig:salt_var}) \cite{thesis}.

\begin{figure*}[t]
\begin{center}
\includegraphics[angle=0,width=14.cm]{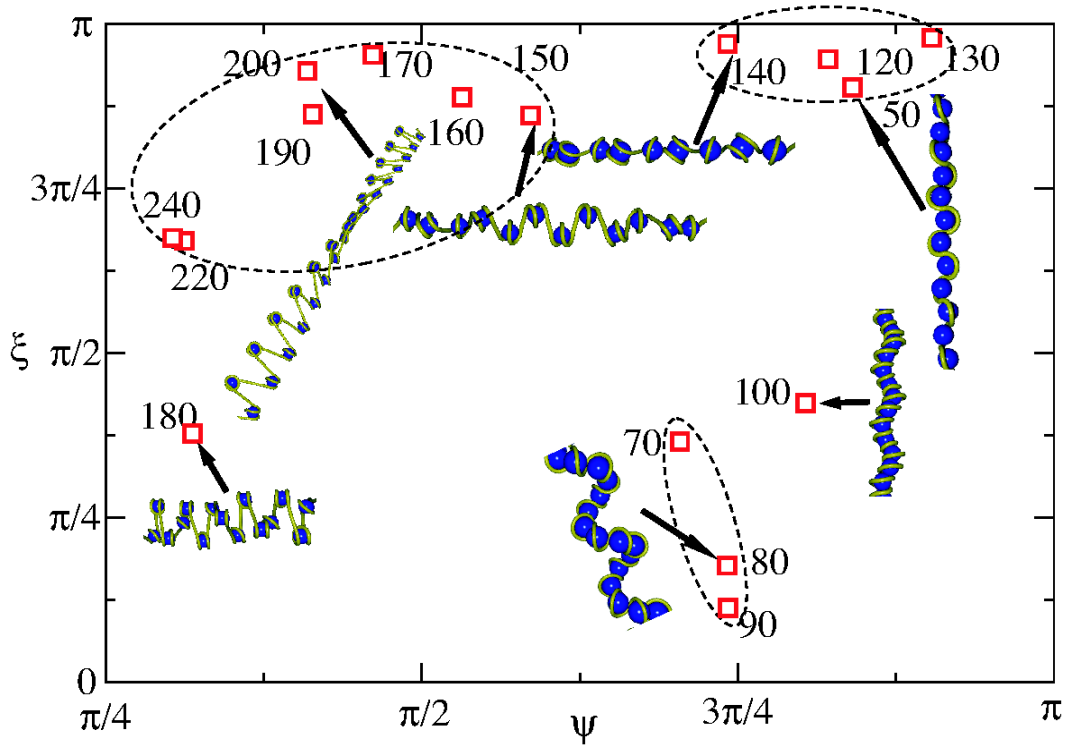}
\smallskip
\caption{\label{fig:chromatin_phase_length}Entry-exit and dihedral  angles diagram 
for various chain lengths per macroion from unconstrained minimization. The corresponding values of $N_{\mathrm{bp}}$ 
are indicated on the graph. The zig-zag region (top left), beads-on-a-string region (top right) and compact 
solenoidal region (bottom right) are shown by closed ellipses. Here $\kappa = 0.8$~nm$^{-1}$ and $Z=15$ are fixed. }
\end{center}
\end{figure*}

\subsection{Two-angle structural  diagram}

In Fig. \ref{fig:chromatin_phase_length}, we have sketched a two-angle diagram in terms of the 
entry-exit angle, $\psi$, and the dihedral angle, $\xi$, summarizing the structural changes of
the fiber with the chain length per unit cell, $N_{\mathrm{bp}} $ (indicated by numbers on the graph). 
As seen, the optimal configurations are mostly gathered in the boundaries of the graph (unlike the behavior found
for varying salt concentration in Fig. \ref{fig:chromatin_phase_kappa}).

The global features may be summarized as follows. For relatively large chain length per unit cell, 
the overall structure changes weakly with the length and exhibits, for the parameters chosen
here, the zig-zag pattern (upper left corner with large dihedral angles 
typically $\xi>\pi/2 $ and small to intermediate entry-exit angles $\pi/4<\psi  <5\pi/8$).  
This is because the structure of the core particle 
is nearly independent of  the chain length in this regime. Note that for 
 $N_{\mathrm{bp}} > 200$, almost 120 base pairs of the PE chain are adsorbed 
 on each sphere, and the rest serves as the linker chain. 
As the chain length decreases down to the length  adsorbed  
on the sphere, i.e., $N_{\mathrm{bp}} = 120$, the diameter of the zig-zag fiber reduces to that of 
the spheres displaying a compact 10~nm fiber with a beads-on-a-string structure (upper right corner
with $\psi > 3\pi/4$ and $\xi>3\pi/4$). 
For smaller chain lengths, there is practically no linker chain present and 
 the core particles will thus be located very close to each other.  
 There is a narrow range of chain lengths $50<N_{\mathrm{bp}}<100$, where the fiber 
adopts a compact solenoidal form, with $0<\xi<\pi/2$ and $\psi$ being distributed around $3\pi/4$. 
Note that the transition between zig-zag and beads-on-a-string structures involves 
an increasing entry-exit angle $\psi$ which maximizes the distance between adjacent spheres
(minimizing their repulsion), while $\xi$ changes only weakly. 
In contrast, the transition between beads-on-a-string and compact solenoidal structures 
involves larger variations in $\xi$. 
The isolated zig-zag pattern with chain length $N_{\mathrm{bp}}=180$ corresponds to 
a local minimum in the binding energy of the fiber as will be discussed below.

\begin{figure*}[!t]
\begin{center}
\includegraphics[angle=0,width=11.cm]{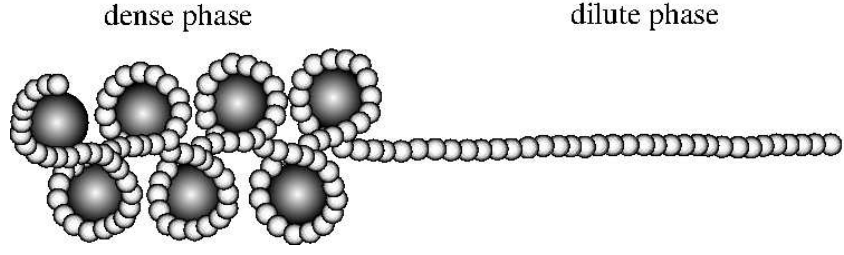}
\smallskip
\caption{\label{fig:ensemble}Schematic view of the ensemble A with fixed number of macroions
adsorbed on a fixed length of the PE chain, exhibiting phase-coexistence between a dense and a dilute phase. }
\end{center}
\end{figure*}

\begin{figure*}[!t]
\begin{center}
\includegraphics[angle=0,width=16.cm]{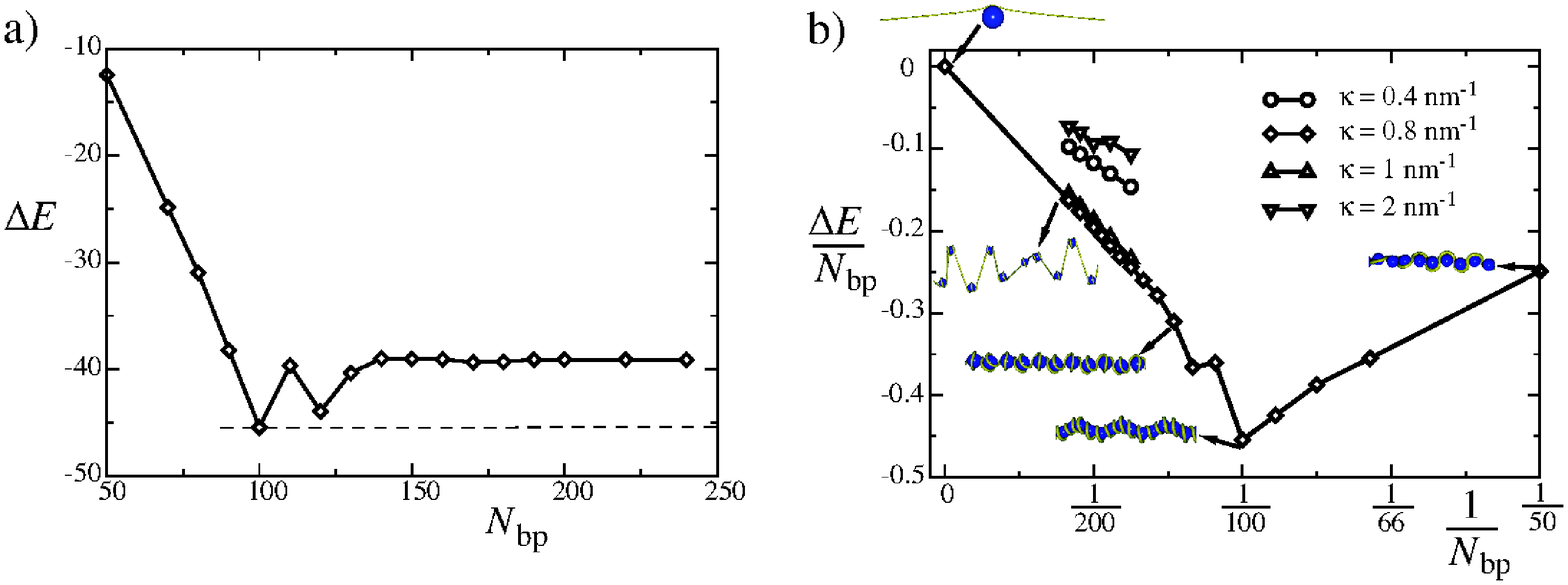}
\smallskip
\caption{\label{fig:ener_length}a) Binding energy of  the complex fiber (per unit cell and in units of $k_{\mathrm{B}}T$) 
as a function of the chain length per unit cell $N_{\mathrm{bp}}$ and b) the binding energy per chain unit length  
as a function of the concentration of adsorbed macroions on the chain, $1/N_{\mathrm{bp}}$, for different values of $\kappa$
as indicated in the graph.}
\end{center}
\end{figure*}

\subsection{Energetic evidence for intra-fiber phase separation}

In the preceding Section, we discussed configurational variations associated with
changes of the  length of the PE segment per unit cell. Now we turn to thermodynamics
and assume that we have a solution of  an infinitely
long poly\-electro\-lyte and oppositely charged 
spherical macroions.  In this case, an interesting question is one of the equilibrium state  of the system. In other words, 
what would be the resultant linear number density of macroions attached to the poly\-electro\-lyte chain and 
the fiber structure? Clearly, a complete 
answer to this question requires a full thermodynamic study by considering the free energy 
in an appropriate ensemble. But within the ground-state-dominance approximation, one can
address this question {\em on the energetic level}, i.e., neglecting entropic effects. 

In order to investigate the stability of the system, we calculate
the fiber binding energy,
${\Delta E} = {\mathcal H}-{\mathcal H}_0$, 
where the first term is the energy of the complex fiber, defined in 
Eq. (\ref{eq:H_fiber}), and the second
term is the energy of a free straight PE chain in the absence of attached spheres, which is used as reference.
Note that  ${\Delta E}$, ${\mathcal H}$ and ${\mathcal H}_0$ are energies per unit cell and in units of $k_{\mathrm{B}}T$.
One can then envision two different ensembles: An ensemble  consisting of a {\em fixed} number of macroions adsorbed
on a {\em fixed} large length of the PE chain (ensemble A), and an ensemble 
consisting of macroions adsorbed from a bulk reservoir (with a fixed chemical
potential) on a {\em fixed} large length of the PE chain (ensemble B). 

Ensemble A is realized in a situation where the total number of macroions is finite and thus  
macroions  are depleted from the solution, and where the total length
of DNA available for adsorption is fixed such that  the 
mean linear concentration of adsorbed macroions is finite and can be treated as an input parameter.
To study thermodynamic equilibrium, in
this case one has to  consider the binding energy per unit cell or sphere 
as a function of the chain length per unit cell, $N_{\mathrm{bp}}$. 
This is shown in Fig. \ref{fig:ener_length}a for fixed $\kappa = 0.8$~nm$^{-1}$ and $Z = 15$. 
As seen,  ${\Delta E}$ appears to have two minima at intermediate values of $N_{\mathrm{bp}}$, i.e., one at about $N_{\mathrm{bp}} = 100$, 
where the fiber exhibits a compact solenoidal structure, and the other 
minimum at about $N_{\mathrm{bp}} = 120$, where a compact 
beads-on-a-string structure is obtained as discussed before. 
Also ${\Delta E}$ shows an overall non-convex behavior between $N_{\mathrm{bp}} = 100$ 
and the limit of an infinitely dilute fiber (with no macroion adsorbed), i.e., $N_{\mathrm{bp}} \rightarrow \infty$.
Thermodynamic stability follows from the standard  common-tangent construction
 (indicated by the broken line in Fig. \ref{fig:ener_length}a),
which indicates a  gas-liquid-type  ``phase coexistence" along the fiber
between a liquid phase of high sphere concentration and a gas phase of vanishing sphere concentration (see Fig. \ref{fig:ensemble}).
In between, the energy has a non-monotonic behavior with values higher than 
the common-tangent line, which thus represents a region of meta-stable structures. 
Adding the one-dimensional translational entropy of adsorbed macroions would lead
to a dilute gas phase with a finite sphere density.
Note that we do not consider the possibility of a higher density of spheres on the chain (i.e. a smaller number $N_{\mathrm{bp}}$ of base pairs 
per unit cell) than corresponding to the minimum in energy in Fig. \ref{fig:ener_length}a.


In the case of ensemble B, one deals with a complex fiber in equilibrium with a reservoir 
of macroions, where the density of macroions on the chain is controlled by 
 the chemical potential of macroions in bulk solution.
 To study thermodynamic equilibrium in this case, one has to consider
the binding energy per unit length of the PE chain ${\Delta E}/{N_{\mathrm{bp}}} = {\mathcal H}/{N_{\mathrm{bp}}}-{\mathcal H}_0/{N_{\mathrm{bp}}}$
 as a function of  the adsorbed macroion density along the chain, $1/N_{\mathrm{bp}}$, 
 which is shown in Fig. \ref{fig:ener_length}b for fixed $\kappa = 0.8$~nm$^{-1}$ and $Z = 15$. We also 
show data  for $\kappa = 0.4$, $1$ and $2$~nm$^{-1}$ for large $N_{\mathrm{bp}}$. Note that the data 
corresponding to $\kappa = 0.8$nm$^{-1}$ lie below the other data, i.e., for this range of 
$N_{\mathrm{bp}}$, the energy of optimal fiber is minimal at $\kappa = 0.8$nm$^{-1}$ as compared
to other values of $\kappa$ \cite{thesis}. 
For the data set for  $\kappa = 0.8$~nm$^{-1}$, ${\Delta E}/{N_{\mathrm{bp}}}$ exhibits 
two minima at intermediate values of macroion density, i.e., one at about  $1/N_{\mathrm{bp}} = 0.01$
($N_{\mathrm{bp}} = 100$),  and the other 
minimum at about $1/N_{\mathrm{bp}} = 0.008$ ($N_{\mathrm{bp}} = 120$)
Note that the location of the minima of ${\Delta E}/{N_{\mathrm{bp}}}$ are in principle different
from the minima  of ${\Delta E}$, but the relative shift is rather negligible.
Phase coexistence between two fiber phases of different densities in this ensemble would 
correspond in fact to three phase coexistence, i.e. matching chemical potentials in the two 
fiber phases and in the bulk macroion solution, and will thus be rather unlikely
since it only occurs when the bulk chemical potential exactly matches the chemical potentials of the dense 
and dilute phases simultaneously. In most cases, the equilibrium  will thus consist of a single phase with a macroion 
density corresponding to the minimum of the energy density in Fig. \ref{fig:ener_length}b.
If the number of spheres in solution is less than needed to homogeneously populate the chain,
one crosses over to ensemble A with a fixed sphere density which is less than the optimal state of ensemble B.

\section{Conclusion and discussion}

In this paper, we present a systematic numerical approach for investigating the structural properties
and energetic stability of complex fibers formed by complexation of a long, semi-flexible poly\-electro\-lyte chain
with many oppositely charged spheres. The complex fiber is described using a chain-sphere cell model, 
in which the detailed structure of the PE chain (locally wrapped around individual spheres) 
as well as interactions between various components (chain segments and spheres) are taken into account. These
interactions include salt-screened electrostatic interactions  between all charged
units (using linear Debye-H\"uckel potential), the bending elasticity of the PE chain as well as the excluded-volume interactions between the 
PE chain and spheres. Here we choose parameters consistent with the DNA-histone system (e.g., using
 the DNA charge and persistence length and a hard-core sphere radius of 5~nm), 
but vary the  salt concentration, sphere charge, and chain length per sphere as control parameters. 

We focus on the optimal structure of the fiber, i.e., the structure which minimizes the effective Hamiltonian 
of the system for a given set of parameters. This amounts to a ground-state analysis, which
is expected to be valid for strongly coupled complexes (large polymer adsorption energy and  relatively 
small thermal fluctuations) \cite{Kunze2,hoda1}. 
Two different schemes are used here for the optimization
procedure: i) unconstrained optimization, and ii) constrained optimization. 
In the former case, we treat the positions of all chain beads within a fiber unit cell as degrees of freedom in addition to 
an independent degree of freedom that defines rotational orientation of two adjacent unit cells (core particles)
with respect to each other. This case therefore requires  a many-variable optimization analysis. 
In the other case, the conformation of the  PE chain on each macroion sphere is fixed according to the optimal 
configuration of a single isolated PE-macroion complex. 
Two such complexes are then linked one by one to form a fiber with
the help of straight linker chains and the only degree of freedom in this case is the relative rotation of 
unit cells  around the linker chain. 
In both cases, a variety of  helical structures are obtained for the fiber including, in particular, the zig-zag patterns. 
There are however qualitative differences between the two models.

In unconstrained minimization,  the PE conformation is free to change on the sphere, 
i.e., upon changing the salt concentration or the macroion charge, the PE chain in each core complex 
may become increasingly  more wrapped or dewrapped.  
As a result, the entry-exit angle of the PE strand  varies accordingly, leading
to dramatic changes in the overall fiber structure. As shown, the optimal fiber structure
shows beads-on-a-string pattern or thick and loose solenoidal patterns at low salt.
At about the inverse screening length $\kappa=0.6$~nm$^{-1}$, the chain 
is already wrapped around the sphere for almost a complete turn. In this case, the spheres are aligned
and form a fiber of diameter about 10~nm, which to some extent resembles the swollen 10~nm beads-on-a-string chromatin fiber.  
For salt concentrations within the physiological regime, zig-zag patterns are found as optimal
structures and the fiber diameter is about 30~nm;  this diameter and the fact that the fiber structure 
is zig-zag in  the physiological regime thus appear to agree with those proposed for chromatin in Refs. 
\cite{Thoma,Allan,Gerchman,Woodcock,Bednar,Horowitz},
although the present model predicts a qualitatively different core particle structure. This is because the linker histone (absent in the present model)
plays a key role in the compaction of chromatin into a 30~nm fiber by gluing together the entering and exiting strands of the DNA
and thus sealing off  the internal structure of the DNA in the nucleosome core particle. In our unconstrained model, the entering and exiting strands of the chain
are displaced and divergent and thus, the density of spheres along the fiber appears to be much smaller (i.e., about one sphere per 11~nm along the fiber axis)
than the value proposed in the cross-linker models (about 5-6 spheres per 11~nm \cite{Schiessel_rev,Thoma,Gerchman,Woodcock,Bednar,Horowitz}). 
For higher salt concentrations, the chain wrapping degree still increases and the
spheres become highly packed along the fiber backbone (for fixed chain length per sphere),
leading to compact solenoidal structures.

 We have also studied the role of sphere charge and chain length
per sphere and their influence on the fiber structure. 
 An important result is that when the chain (linker) length per macroion is treated as a free parameter,  
the binding energy (per sphere or unit cell) is found to take a  non-convex shape at intermediate salt concentrations 
when plotted as a function of
the chain length per sphere. A simple common-tangent construction indicates a  gas-liquid-type ``phase coexistence" along the fiber,
i.e., a part of the PE chain forms a dense complex with macroions phase-separating from 
a dilute phase along the fiber.

In the constrained optimization model, the core particle structure is fixed 
and only the relative rotation of the core particles (unit cells) is allowed in order to minimize the 
fiber Hamiltonian.  As shown, for large sphere charge and within the physiological salt regime, 
the fiber shows 
zig-zag pattern and sphere densities of about 5-6 spheres per 11~nm length along the fiber axis. 
Indeed, both the global structure as well as  the fiber density and diameter in this case resemble more closely the structures
found or proposed for chromatin in Refs. \cite{Thoma,Gerchman,Woodcock,Bednar,Horowitz}. 
Interestingly, the dense zig-zag pattern results from the  fact that the entering and exiting strands form 
a cross pattern leading to a small entry-exit angle; the form of the entry-exit strands in this case
tends to resemble the stem-like form expected in chromatin in the presence of the linker histone 
 \cite{Woodcock,Bednar} and is therefore found to be a similarly important factor in order to obtain dense fiber structures
 in our model. The constrained model is thus similar to the two-angle model proposed by Woodcock et al. \cite{Woodcock}, except that 
here we employ a realistic interaction potential between the core particles. 

We emphasize again that the present results are obtained for a generic charged chain-sphere 
complex fiber  and no direct comparison with the 30~nm chromatin fiber (which involves additional structural 
and specific details) should be attempted. 
Especially, the linker histone effects or other factors such as histone tails or the specific shape of the histone octamer  
are not incorporated within the present model, and the constrained optimization model is only a simple way to mimic the 
linker-histone stabilization of the nucleosome. On the other hand, 
our model and its predictions are applicable to the mixtures of DNA and  
synthetic oppositely charged  spheres and suggest that chromatin-like structures should also be observable in such systems. 
The main advantage of the present approach is that it accounts for the detailed wrapping state of the PE chain within
a fiber as well as the essential intra-fiber interactions, and that it can easily be generalized to more sophisticated models.  

In this work, we have also neglected the excluded-volume interaction between chain segments. (Note that the chain-sphere
volume interaction and the finite radius of DNA is taken into account via the hard-core radius of the sphere as 
explained in the text.) The chain-chain excluded-volume interaction is expected to play a role
only at high salt concentration (typically beyond 100mM monovalent salt for intermediate sphere charge), where the chain 
wraps around the spheres more than two turns. (For a complete analysis and possible improvements of the basic
chain-sphere scheme used here see the discussions in Refs. \cite{Kunze2,hoda1,thesis}.) 

Another interesting problem is to study the response of the fiber to an externally applied stress. 
Recent experiments have shown that the response of  chromatin to an external mechanical stress can provide
more insight into the DNA-histone interactions and the detailed conformation of the fiber \cite{Schiessel_rev,Widom,Yao,Cui,Katritch}. 
This problem has been considered in several recent theoretical  studies \cite{Schiessel_chromatin,Schiessel_rev,Schiessel_EPL,Kulic04,Kunze2}. 
Finally, one may also consider an inhomogeneous  fiber within the present model by assuming that the linker length may vary from
one unit cell to the other. In this case,  the fiber may exhibit overall deformations and large-scale conformational fluctuations 
\cite{Woodcock}.  


\begin{acknowledgement}
A.N. is supported by the Royal Society, the Royal Academy of Engineering, and the British Academy.
\end{acknowledgement}


\end{document}